\pdfoutput=1

\documentclass[10pt,twocolumn,letterpaper]{article}

\usepackage[final]{style/cvpr}      

\definecolor{cvprblue}{rgb}{0.21,0.49,0.74}
\usepackage[pagebackref,breaklinks,colorlinks,allcolors=cvprblue]{hyperref}

\usepackage{amsmath,amsfonts,bm}

\def\eqref#1{equation~\ref{#1}}

\def\1{\bm{1}}

\DeclareMathAlphabet{\mathsfit}{\encodingdefault}{\sfdefault}{m}{sl}
\SetMathAlphabet{\mathsfit}{bold}{\encodingdefault}{\sfdefault}{bx}{n}

\usepackage{hyperref}
\usepackage{url}
\usepackage{graphicx}
\usepackage{tcolorbox}
\usepackage{float}
\usepackage{enumitem}
\usepackage{comment}

\usepackage{hyperref}
\usepackage{url}
\usepackage{pifont}
\usepackage{placeins}

\def\paperID{11740} 
\def\confName{CVPR}
\def\confYear{2025}

\title{The Efficacy of Transfer-based No-box Attacks on Image Watermarking:\\ A Pragmatic Analysis}

\author{
  Qilong Wu,  Varun Chandrasekaran\\
  University of Illinois Urbana-Champaign\\
  \{qilong3, varunc\}@illinois.edu
}

\begin{document}
\onecolumn
\maketitle
\begin{abstract}

Watermarking approaches are widely used to identify if images being circulated are authentic or AI-generated. Determining the robustness of image watermarking methods in the ``no-box'' setting, where the attacker is assumed to have no knowledge about the watermarking model, is an interesting problem. Our main finding is that evading the no-box setting is challenging: the success of optimization-based transfer attacks (involving training surrogate models) proposed in prior work~\cite{hu2024transfer} depends on impractical assumptions, including (i) aligning the architecture and training configurations of both the victim and attacker's surrogate watermarking models, as well as (ii) a large number of surrogate models with potentially large computational requirements. Relaxing these assumptions i.e., moving to a more pragmatic threat model results in a failed attack, with an evasion rate at most $21.1\%$. We show that when the configuration is mostly aligned, a simple non-optimization attack we propose, OFT, with one single surrogate model can already exceed the success of optimization-based efforts. Under the same $\ell_\infty$ norm perturbation budget of $0.25$, prior work~\citet{hu2024transfer} is comparable to or worse than OFT in $11$ out of $12$ configurations and has a limited advantage on the remaining one. The code used for all our experiments is available at \url{https://github.com/Ardor-Wu/transfer}.

\end{abstract}
\section{Introduction}
\label{sec:intro}

\begin{figure*}[]
   \centering
   \begin{tabular}{cccc}
       \includegraphics[width=0.125\linewidth]{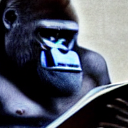} &
       \includegraphics[width=0.125\linewidth]{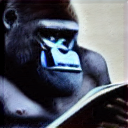} &
       \includegraphics[width=0.125\linewidth]{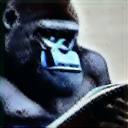} &
       \includegraphics[width=0.125\linewidth]{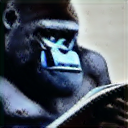} \\
       Unwatermarked & Watermarked & OFT (Unnorm., $k=1$) & OFT (Norm. $k=1$) \\
       \includegraphics[width=0.125\linewidth]{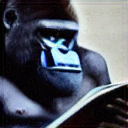} &
       \includegraphics[width=0.125\linewidth]{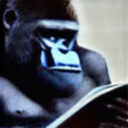} &
       \includegraphics[width=0.125\linewidth]{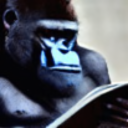} &
       \includegraphics[width=0.125\linewidth]{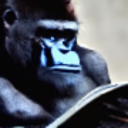} \\
       \citet{hu2024transfer} $k=50$ & DiffPure~\citep{saberi2023robustness} $t=0.1$ & DiffPure~\citep{saberi2023robustness} $t=0.2$ & DiffPure~\citep{saberi2023robustness} $t=0.3$
   \end{tabular}
   \caption{Watermarked images generated by Stable Diffusion and their perturbed versions in different attacks that successfully evade detection. }
   \label{Image Samples}
\end{figure*}

Generative AI has demonstrated great promise in creating high-quality images~\citep{ramesh2021zeroshot, saharia2022imagen,rombach2022high,midjourney2022,balaji2022makeascene,bai2024meissonic}. However, these technologies can also be misused for generating fake, misleading images~\citep{brandom2024} and art plagiarism from style mimicry~\citep{shan2023glaze}. Thus, detecting AI-generated images is of significant interest. The most popular approach~\citep{saharia2022photorealistic,ramesh2021zeroshot,rombach2022high}, image watermarking, holds significant promise for identifying AI-generated images~\citep{tanaka1990digital,zhu2018hidden,zhang2019robust,jia2021mbrs,tancik2020stegastamp,jiang2024certifiably}. In image watermarking, the watermark (secret) is embedded in the image being shared; the image is detected as watermarked when the decoded watermark matches the secret embedded. 

However, attackers may use evasion attacks~\citep{jiang2023evading,hu2024transfer,saberi2023robustness,zhao2023invisible,chen2020hopskipjumpattack,an2024box,liu2024image} to strategically edit the watermarked image to evade detection, i.e., add a crafted perturbation such that the target watermark detector falsely classifies the ``attacked'' image as not watermarked. As a result, the robustness of image watermarking against evasion attacks is crucial in practice. Existing work has demonstrated that image watermarking techniques are not robust in the white-box~\citep{jiang2023evading} or black-box~\citep{jiang2023evading,chen2020hopskipjumpattack,lukas2023leveraging,an2024box} settings, where the attacker can at least query the watermarking detector. However, detector access can be unrealistic in practice~\cite{grosse2024towards}. In contrast, the robustness of image watermarking in the ``no-box'' setting, where the attack has no knowledge about the watermarking model, remains an interesting problem. Regeneration-based attacks, built using strong generative models (such as publicly pre-trained diffusion models~\citep{stablediffusion,dalle2,imagen}) can ``de-noise'' the watermark~\citep{nie2022diffusion,saberi2023robustness,zhao2023invisible,an2024box,liu2024image}. However, such models are not applicable when the generated image is o.o.d w.r.t the diffusion model's training data; training such (customized) models requires expensive computations~\citep{liu2024image}. Transfer-based attacks~\citep{hu2024transfer} utilize surrogate watermarking models to seed their optimization procedure. They are built on the intuition that if surrogate models' decoded messages are flipped by adding perturbations, the victim models' decoded messages can also be disrupted. However, these attacks make strong assumptions about the knowledge of the watermarking method (we call this the {\em alignment} assumption) and computational capabilities afforded to the attacker. We aim to relax these assumptions: {\em in a realistic no-box setting, attackers have no knowledge about the victim watermarking and have limited computational resources.}

When both assumptions are relaxed, unsurprisingly, prior work~\citep{hu2024transfer} fails. We observe that relaxing the alignment assumption leads to the failure of most transfer-based attacks. However, we see gains when relaxing the computational requirements: we propose an inexpensive attack, Optimization-Free Transfer (OFT), that works in settings where computation is reduced. It is on par, or even better than the attack of~\citet{hu2024transfer} and a more powerful regeneration-based attack (Diffusion Purification~\citep{saberi2023robustness}), showing its effectiveness despite its simplicity. 

Our contributions are as follows:

\begin{itemize}
\item We empirically find that the existing transfer-based attack's~\citep{hu2024transfer} success relies on unrealistic assumptions of both aligned configurations between the surrogate models and the victim model and high computation capabilities. Relaxing either assumption results in this attack failing.
\item We propose a simple and inexpensive optimization-free transfer attack (OFT) that is on par or even better compared with~\citet{hu2024transfer} and Diffusion Purification~\citep{saberi2023robustness} when configurations are aligned.
\end{itemize}

\section{Notation \& Background}
\label{sec:notation and background}

Image watermarking is a promising approach to recognizing model-generated images. Many state-of-the-art methods utilize an encoder-decoder structure~\cite{zhu2018hidden, jia2021mbrs, zhang2019robust, tancik2020stegastamp}, where the image is encoded to a lower dimensional representation where the watermark is added before it is decoded back; others~\cite{zhao2023recipe} incorporate watermarking into the image generation model itself. We will first present a general framework for how watermarking is achieved (in the former category), and then describe approaches we consider in this paper.

\subsection{A General Watermarking Framework}
\label{sec:general}

\noindent{\bf Parameters:} The parameters include $w$ (image width), $h$ (image height), $\ell$ (secret message length), 
and $\delta$ (watermark detection threshold).

\vspace{0.3em}
\noindent\underline{\bf Encoder:}
The encoder takes two inputs: $x \in \mathbb{R}^{3 \times w \times h}$, which is the image to be watermarked, and $s \in \{0,1\}^\ell$, which is the secret (needed for watermarking). The image encoder $e(.)$ converts the image to an image embedding, denoted as $x_e = e(x)$. The secret encoder $g(.)$ converts the secret message to a secret embedding, represented as $s_g = g(s)$. The mapping function $h(., .)$ maps the image and secret message to some embedding space, resulting in a projected image with encoded secret, written as $x_h = h(x_e, s_g)$. This is projected back to the image space by function $f$ to obtain the watermarked image i.e., $x_{wm} = f(x_h, x)$. 
The overall encoder for the watermarked image ($x_{wm} \in \mathbb{R}^{3 \times w \times h}$) is represented as $\texttt{Enc}(s, x)$. 

\vspace{0.3em}
\noindent\underline{\bf Decoder:} The decoder $\texttt{Dec}(.)$ produces a decoded secret message, represented as $s' = \texttt{Dec}(x_{wm}) \in \{0,1\}^\ell$. 

\vspace{0.3em}
\noindent\underline{\bf Detector:} The detector takes two inputs: $x_{wm} \in \mathbb{R}^{3 \times w \times h}$, which is the image for watermarking detection, and $s \in \{0,1\}^\ell$, which is the secret for the watermark. Its functionality can be expressed mathematically as:

$\texttt{Det}(s,\texttt{Dec}(x_{wm})) = \begin{cases}
1, & \texttt{if } \texttt{BA}(s,s') \geq 1-\delta \\
0, & \texttt{otherwise}
\end{cases}$

Note that the above is for the one-tailed detection and the two-tailed detection variant is as follows:   
\[
\texttt{Det}(s, \texttt{Dec}(x_{wm})) =
\begin{cases} 
    1, & \text{if } \texttt{BA}(s, s') \geq 1 - \delta \text{ or } \texttt{BA}(s, s') \leq \delta \\
    0, & \text{otherwise}
\end{cases}
\]

\vspace{0.3em}
\noindent\underline{\bf Critic:} Some approaches concurrently train a critic network designed to determine if a given image is watermarked or not. Encoders are optimized to fool the critic.

\vspace{0.3em}
\noindent\underline{\bf Noise Layers:} Noise layers add, for example, Gaussian noise to the watermarked images before they are sent to the detector during training. This is to ensure the robustness of the watermark against common transmission noise. 

\vspace{0.3em}
\noindent\underline{\bf Training Details:} The model is trained to minimize the image distortion due to watermarking, while maximizing detectability $\texttt{BA}(s,\texttt{Dec}(\texttt{Enc}(x,s)))$, where \texttt{BA} stands for bit-wise accuracy. If the image is watermarked, then $\texttt{BA}(s,s')$ will be close to $1$. If not, then the number of matched bits is expected to be of binomial distribution $\ell \times \texttt{BA}(s,s') \sim \mathbb{B}(\ell,\frac{1}{2})$. So, the model can verify if a given image is watermarked by measuring the bit-wise accuracy of the recovered secret.

\subsection{Watermarking Methods}
\label{sec:examples}

Different watermarking methods have various architectures, particularly in their encoder designs under our framework. In our work, we consider 4 salient approaches: \texttt{HiDDeN, MBRS, RivaGAN} and \texttt{StegaStamp}. We pick \texttt{MBRS, RivaGAN, StegaStamp} following~\citet{saberi2023robustness}, and \texttt{HiDDeN} as it is the first representative learning-based watermark method. We outline the architectural details of their implementations, and present other salient features we use in experiments for target models in Table~\ref{tab:watermark-comparison}. Full mathematical formulations of encoders, decoders, etc. can be found in Appendix~\ref{sec:appendix_wm}, along with other relevant details.

\begin{table*}[t]
\centering
\caption{{\bf Comparison of Different Watermarking Methods:} Observe the similarity between \texttt{HiDDeN} and \texttt{StegaStamp}'s encoders. 
}
\label{tab:watermark-comparison}
\begin{tabular}{l p{3.8cm} p{3.5cm} p{2cm} p{3cm}}
\toprule
{\bf Method} & {\bf Encoder} & {\bf Decoder} & {\bf $\ell$ (bits)} & {\bf Training Dataset} \\
\midrule
\texttt{HiDDeN}~\cite{zhu2018hidden} & 5/8 Conv-BN-ReLU blocks and one Conv layer & Conv layers, average pooling, linear layer & 20/30/64 & DiffusionDB~\citep{wang2022diffusiondb} or MidJourney~\citep{iulia_turc_gaurav_nemade_2022} \\

\vspace{2.5mm}

\texttt{MBRS}~\cite{jia2021mbrs} & Conv-BN-ReLU, ExpandNet, SENet blocks & Conv-BN-ReLU, SENet blocks, Conv layer, reshaping & 64 & DALLE-2~\citep{dalle_gallery_2023} \\

\vspace{2.5mm}

\texttt{RivaGAN}~\cite{zhang2019robust} & Conv3D-Tanh-BN3D blocks & Conv3D-Tanh-BN3D blocks & 32 & Hollywood 2~\citep{marszalek2009actions} \\

\vspace{2.5mm}

\texttt{StegaStamp}~\cite{tancik2020stegastamp} & U-Net style architecture composed of Conv-ReLU blocks & Spatial transformer, Conv and dense layers, sigmoid & 100 & MIRFLICKR~\citep{huiskes08} \\
\bottomrule
\end{tabular}
\label{comparison}
\end{table*}

\section{Robustness of Watermarking Schemes}
\label{sec:robustness}

\noindent{\bf How is robustness evaluated?} In the status quo, robustness is measured in the worst-case, by ``perturbing'' the watermarked image so that detection fails; this means that the secret encoded is incorrectly extracted. This can be formalized as an {\em evasion attack}, where adversarial noise is added to watermarked images. In contrast, {\em spoofing attacks} try to edit unwatermarked images such that the resulting images are falsely detected as watermarked. This can be used to ruin the watermarking model's reputation, but is not within the scope of our paper.

\vspace{1mm}
\noindent{\bf In what settings are evasion attacks performed?} There are several settings for evasion attacks on image watermarking: white-box~\citep{jiang2023evading}, black-box~\citep{jiang2023evading,chen2020hopskipjumpattack,lukas2023leveraging,an2024box}, and no-box. From an attacker's perspective, the aforementioned ordering is from easy to hard. We focus more on the no-box setting as it is assumed to be the most realistic. More details about the other two settings are presented in Appendix~\ref{app:attack_models}.

\vspace{1mm}
\noindent{\bf What is the no-box setting?} It is the most challenging setting as the attacker has no knowledge about the detector, and cannot access the detector (even through API calls). We also assume the attacker does not know the prompts used to generate watermarked images (if they are generated by text-to-image models). Evasion attacks in this setting remain challenging. 

\subsection{A General Attack Framework}
\label{attack_framework}

The attacker's objective is to design an algorithm \texttt{Attk} that 
takes as input watermarked images $x_{wm}$ (which are watermarked by the ``victim'' model with some unknown method). Most attack methods assume target watermarked images come in batches, while the attack is more challenging if they come individually. This algorithm also utilizes some auxiliary information, which we will detail soon. More formally, 
\[
x_{a} = \texttt{Attk}(x_{wm}, \text{aux})
\]

\noindent The algorithm must meet three properties:

\begin{enumerate}
    \item \underline{\em Ensure successful evasion:}
    \[
        \texttt{Det}(\texttt{Dec}(x_{wm}), \texttt{Dec}(x_a)) = 0
    \]

    \item \underline{\em Minimize distortion:}
    \[
        \texttt{Dis}(x_{a}, x_{wm}) \text{ is low}
    \]
    where $\texttt{Dis}(\cdot, \cdot)$ represents a distortion function. For example, it can be represented as:
    \[
        \| x_{a} - x_{wm} \|_{\infty} \leq r
    \]
    with $r$ being a fixed $\ell_\infty$ budget.

    \item \underline{\em Be inexpensive:} A practical attack should keep the computation cost of $\texttt{Attk}(\cdot)$ low.
\end{enumerate}

\vspace{1.5mm}

\noindent Depending on the nature of the attack, the attacker may use the following auxiliary (aux) information: 
\begin{itemize}
    \item A set of $k$ surrogate encoders $\texttt{Enc}_i(\cdot)$, $i \in [k]$ and decoders $\texttt{Dec}_i(\cdot)$, $i \in [k]$ (where $[k] = \{1,\cdots,k\}$). Note that these encoders/decoders {\em need not} be the same as that of the victim watermarking method.
    \item A generative model $\mathcal{A}$ and projection function $\phi$ s.t. $\texttt{Dis}(x, \mathcal{A}(\phi(x)))$ is low. 
\end{itemize}

\subsection{Examplar No-Box Attacks}
\label{subsec:no_box_examples}

There are several other no-box attacks that are variants of the ones mentioned below, but not as effective; they are discussed in detail in Appendix~\ref{app:attack_models}.

\vspace{1mm}
\noindent{\bf Regeneration-based Attacks:} Here, the attacker aims to project the image, perturb it, and then regenerate the ``noisy'' version. Broadly, this can be done in three ways:
\begin{itemize}
\item \underline{\em Naive regeneration (NR)} i.e., $x_a = \mathcal{A}(\phi(x_{wm}))$, as done by~\citet{an2024box} 
\item  \underline{\em Noise-then-embed (NTE)} i.e., $x_a = \mathcal{A}(\phi(x_{wm} + \eta))$, where $\eta \sim \mathcal{N}(0, \sigma^2I)$, as done by~\citet{saberi2023robustness,nie2022diffusion,zhao2023invisible,an2024box} 
\item  \underline{\em Embed-then-noise (ETN)} i.e., $x_a = \mathcal{A}(\phi(x_{wm}) + \eta)$, where $\eta \sim \mathcal{N}(0, \sigma^2I)$, as done by~\citet{liu2024image,zhao2023invisible} 
\end{itemize}
\smallskip
Different approaches can be formalized by the exact choice of $\mathcal{A}$ and $\phi$, with most promising results shown when $\mathcal{A}$ is a pre-trained diffusion model. While $\phi$ can be replaced by $\texttt{Enc}(.)$ used for watermarking, most approaches make no assumptions about $\phi$, making these attacks no-box. More details about the these approaches are found in Appendix~\ref{app:attack_models}.

\vspace{1.5mm}
\noindent{\bf Transfer-based Attacks:} Proposed by~\citet{hu2024transfer}, these attacks build atop the intuition that if surrogate models' decoded messages are flipped by adding perturbations, the victim model's decoded message can also be disrupted. Formally, they solve the following optimization problem:
\begin{align*}
& \min_\varepsilon \frac{1}{k} \sum_{i=1}^k \texttt{Dis}(\texttt{Dec}_i(x_{wm} + \varepsilon), w_i^t)) \\
& \text{s.t.} \quad ||\varepsilon||_{\infty} \leq r \\
& \frac{1}{k} \sum_{i=1}^k \texttt{BA}(\texttt{Dec}_i(x_{wm} + \varepsilon), w_i^t)) > 1 - \gamma
\end{align*}
where:
\begin{itemize}
\item $w_i^t = \overline{\texttt{Dec}_i(x_{\text{wm}})}$, where $\overline{\texttt{Dec}_i(x_{\text{wm}})}$ is the bit-wise flip of $\texttt{Dec}_i(x_{\text{wm}})$.
\item $\texttt{Dis}$ is set to be the mean square error.
    \item $r$ is the perturbation budget.
    \item $\gamma$ is the sensitivity hyperparameter.
\end{itemize}

\subsection{But Are These Attacks Practical?}

While the results from all of these attacks (with the exception of NR) are impressive, the regeneration-based attacks make the least assumptions about the nature of the watermarking method. To evaluate this, we analyzed the code and formulation of all the regeneration-based attacks we discuss earlier, and notice if there are any dependencies between $\texttt{Attk}$ and the watermarking process (in terms of shared architecture, knowledge, dataset etc.). {\em We could not find any dependencies.} Thus, for the remainder of the paper, our focus will be solely on transfer-based attacks: while not obvious from the equations above, current transfer attacks are not as general as formulated; as we will show in \S~\ref{sec:results} (and in Figure 6 of their draft\footnote{\url{https://arxiv.org/abs/2403.15365}}), they have a strong dependence on ``aligning'' the surrogate and victim models. 

\smallskip
\noindent{\bf What is alignment?} This is the process followed to orient the ``configurations'' of the surrogate model to the victim model. This implicitly requires the knowledge of the victim model, violating the no-box setting. The configuration includes, but is not limited to, watermarking protocols, model architecture, secret length, dataset, input dimension, and types of noise added during training. The implementation of existing transfer-based attack~\citep{hu2024transfer} aligns most of the above, with the exception of choice of secret and dataset, while the architecture difference is zero or minimal at best. 

In conclusion, transfer-based attacks in the status quo make two strong assumptions:

\vspace{0.5mm}
\noindent \underline{\em Assumption 1: Large compute capabilities.} The attacker has computational capabilities to launch sophisticated attacks. This includes capabilities to train dozens of surrogate models.

\vspace{0.5mm}
\noindent \underline{\em Assumption 2: Extensive prior knowledge (or the ``alignment'' assumption).} The attacker {\em knows the watermarking method and most architectural information of the victim watermarking model}.

\smallskip
\noindent{\bf Why are these assumptions ``impractical''?} Most studies of adversarial robustness focus on {\em worst-case guarantees}, when the attacker is determined and aware of the defensive countermeasures. Such studies assume that the attacker is presented with information about the training data, architecture of the countermeasures, and is provided with extensive computational support to attack these systems. But what about {\em average-case guarantees?} Most practical deployments of watermarking technologies are truly opaque. While organizations may share details about their deployments~\citep{deepmind_synthid_2023}, there is limited evidence to show that organizations productionize what they share. Even if they do, there is virtually no guarantee that the attacker has information about how these watermarking models are trained, or what architecture is used, unless the attacker has insider information. Other organizations share no information about their watermarking practices at all~\citep{openai_watermarking_2024}. In their seminal work,~\citet{apruzzese2023real} interview various organizations, and state that {\em ``it is sensible to conclude that an attacker would opt for these [simpler] methods over more computationally expensive ones''}, implying that costs are a big factor in the attacker's mind. This observation is echoed in the work of~\citet{grosse2024towards}, which clearly states that {\em ``research is often too generous with the attacker, assuming access to information not frequently available in real-world settings''}. What makes the problem more complicated for the attacker is the fact that ML components (like watermarking) are but a small piece in a larger system that they would need to infiltrate~\cite{bieringer2022industrial,mink2023security}. While the no-box setting is assumed to be one of the more practical threat models, our work aims to understand its practicality when the aforementioned assumptions are relaxed; this will provide a better estimate of the average-case efficacy of such transfer-based attacks.

\section{An Inexpensive, Optimization-Free Attack}
\label{sec:simple}

Recall that prior transfer-based attacks require the attacker to utilize expensive constraint optimization-based approaches to design the ``fake'' watermark to introduce a new ``perceived secret'' (i.e., Assumption 1). We want to understand if this cost is necessary, or if it can be reduced. Our intuition is that if the perceived secret is (bit-wise) flipped for the surrogate model after adding a perturbation $\varepsilon$, then the perceived secret of the target model may also be flipped, leading to evasion. One candidate perturbation can be the watermark from a surrogate model, whose secret is a flipped version of what it perceives. A slightly more advanced version is the aggregated version of watermarks with such flipped secrets from multiple surrogate models. To minimize perceptual degradation, normalization may be used. We outline the steps for our {\em Optimization-Free Transfer (OFT)} attack below:

\vspace{1mm}
\noindent{\bf Step 1. Decode and Flip Perceived Secret:} For each surrogate model decoder $\texttt{Dec}_i$, we first decode the secret message for each of them and take the bit-wise flip.

\[
s_i=\overline{\texttt{Dec}_i(x_{wm})}\
\]

\vspace{1mm}
\noindent{\bf Step 2. Compute Individual Perturbations:} For each surrogate encoder, we calculate:

\[
\varepsilon_i = \texttt{Enc}_i\left(s_i,\, x_{wm}\,\right) - x_{wm},
\]

\vspace{1mm}
\noindent{\bf Step 3. Aggregate Perturbations:} We combine the individual perturbations using the aggregation function \( \texttt{Aggr} \):

\[
\varepsilon = \texttt{Aggr}\left(\varepsilon_1,\, \varepsilon_2,\, \dotsc,\, \varepsilon_k\right).
\]

The aggregation function \( \texttt{Aggr} \) can be the pixel-wise mean or median.

\vspace{1mm}
\noindent{\bf Step 4. Normalize Perturbation (if Necessary):} If normalization is required to adhere to the attack budget, say $\|\varepsilon\|_{\infty} \leq r$, we adjust \( \varepsilon \) by clamping its values so that each element lies within \(-r\) and \( r \):

\[
\varepsilon_i = \begin{cases}
    -r, & \text{if } \varepsilon_i < -r, \\
    \varepsilon_i, & \text{if } -r \leq \varepsilon_i \leq r, \\
    r, & \text{if } \varepsilon_i > r.
\end{cases}
\]

This ensures that the perturbation's individual components do not exceed the attack budget \( r \) in magnitude. Observe that we take the same \( \ell_\infty \) norm as in~\citet{hu2024transfer}, but other norms can also be enforced.

\vspace{1mm}
\noindent{\bf Step 5. Generate Attacked Image:} Finally, we apply the aggregated perturbation to the original watermarked image:

\[
x_{a} = x_{wm} + \varepsilon
\]

\section{Relaxing Both Assumptions}
\label{sec:setup}

\noindent{\bf Setting 1:} Relaxing the first assumption implies that the attacker uses our OFT attack as opposed to the one by~\citet{hu2024transfer}. A special case of this assumption is also reducing the number ($k$) of surrogate models. To further highlight the efficacy of our attack, we compare it with a powerful regeneration-based attack, diffusion purification (DiffPure)~\citep{saberi2023robustness}, as well. DiffPure first adds creates a noisy version of $x_{wm}$ as $x_t\sim \mathcal{N} \left( \sqrt{\bar{\alpha}_t} x_{wm}, (1 - \bar{\alpha}_t) I \right)
$, where $\bar{\alpha}_t$ is an increasing function of $t$ that spans from $1$ to $0$ as $t$ progresses from $0$ to $1$; $t$ is the attack strength. Then the attacked image is given by $x_a = \mathcal{A}(\phi(x_t))$, where $\mathcal{A}$ is the denoiser, and $\phi$ is a diffusion model, making $\mathcal{A}(\phi(.))$ a guided-diffusion. 

\vspace{1mm}
\noindent{\bf Setting 2:} Relaxing the second assumption implies that the attacker (a) is unaware of the architecture of the victim watermarking technique, and (b) makes no effort to align them. To this end, we test~\citet{hu2024transfer}'s attack on four different target watermarking methods, \texttt{MBRS}, \texttt{RivaGAN}, \texttt{StegaStamp}, and \texttt{HiDDeN}. The surrogate models' watermarking method is \texttt{HiDDeN}.

\vspace{1mm}
\noindent{\bf Note:} All our experiments are conducted on a single NVIDIA H100 GPU.

\subsection{Implementation Details}

\subsubsection{Models}
\label{ss:models}

\noindent{\bf Surrogate Models:} We use the checkpoints shared by authors of~\citet{hu2024transfer}. More information is available in \S~6.1 (Experiment Setup) in~\citet{hu2024transfer}.

\vspace{1mm}
\noindent{\bf Target Models:} In setting 2(a), the target models are obtained as follows:
\begin{enumerate}
\item \texttt{StegaStamp}: official checkpoint released.\footnote{\url{https://watermarkattack.s3.amazonaws.com/stegastamp_pretrained.zip}}
\item \texttt{RivaGAN}: from invisible-watermark package.\footnote{\url{https://github.com/ShieldMnt/invisible-watermark}} 
\item \texttt{MBRS}: from the repository of~\citet{saberi2023robustness}\footnote{\url{https://github.com/mehrdadsaberi/watermark_robustness/tree/main/MBRS_repo}}. From Table~\ref{tab:watermark-comparison}, we can see that this method is, at a high-level, architecturally similar to \texttt{HiDDeN}. Thus, as a control, we align the configuration as best as we could. More details are in Appendix~\ref{sec:app_align}. Thus, there are 2 versions of this watermark, and we report results only for the ``aligned'' \texttt{MBRS}.
\item \texttt{HiDDeN}: checkpoint shared by~\citet{hu2024transfer}. More information is available in \S~6.1 (Experiment Setup) of their work.
\end{enumerate}
The detailed configuration of different target models can be found in Appendix~\ref{sec:appendix_wm}.

\vspace{1mm}
\noindent{\bf Watermarking Configurations:} In setting 2(b), we test on (a) secrets of length $\ell \in \{20, 30, 64\}$, (b) CNN and ResNet architecture described in~\citet{hu2024transfer} (details soon follow), and (c) the target model trained on DiffusionDB~\citep{wang2022diffusiondb} or MidJourney~\citep{iulia_turc_gaurav_nemade_2022} datasets. Specifically, (1) for the CNN architecture, the encoder consists of 4 Conv-BN-ReLU blocks, while the decoder consists of 7 Conv-BN-ReLU blocks, and (2) for ResNet architecture, the encoder consists of 7 Conv-BN-ReLU blocks, and the decoder is the ResNet-18~\cite{he2016deep}. Observe that the difference in the encoder architecture between CNN and ResNet architecture is limited: they only differ in the number of blocks. The main difference is the decoder. For the dataset, the same processing is followed as in~\citet{hu2024transfer} (details in \S~6.1 of their paper). We evaluate $10$ different surrogate model checkpoints to factor the randomness from different checkpoints.

For the surrogate models with $\ell=30$ and CNN architecture, we use the model shared by~\citet{hu2024transfer}. For the remaining, we train them ourselves using~\citet{hu2024transfer}'s implementation. Note that we tune the hyperparameters for CNN architectures because the ones provided by~\citet{hu2024transfer} do not reach a reasonable \texttt{BA}. \footnote{We perform grid search on \texttt{encoder\_loss} weight in $\{0.7, 0.35, 0.175, 0.0875\}$ and \texttt{decoder\_loss} weight in $\{1, 2, 4, 8\}$. The best configuration was obtained after we choose \texttt{encoder\_loss} weight $0.7$ and decoder\_loss weight $4$. More details are in Appendix~\ref{64bits_cnn_tuning}.} 

\subsubsection{Evaluation Metrics}

\noindent{\bf 1. Evasion Rate:} This is the proportion of the perturbed images that evade the target watermark-based detector. 
\[
\text{Evasion rate} = \frac{\| \{ x_{a} \mid \texttt{Det}(s, \texttt{Dec}(x_{a})) = 0 \}\|}{\| \{ x_a \}\|} 
\]

We use evasion rate for the same false positive threshold setting criteria as in~\citet{hu2024transfer} ($\leq 10^{-4}$). 

\vspace{1mm}
\noindent{\bf 2. Average \texttt{BA}:}
\[
\text{Average }\texttt{BA} = \frac{1}{N} \sum_{i=1}^{N} \texttt{BA}(x_{a_i})
\]

where \( N \) is the total number of attacked images, and \( \text{BA}(x_{a_i}) \) represents the bit-wise accuracy of the attacked image \( x_{a_i} \). 

\vspace{1mm}
\noindent{\bf 3. \# Matched Bits:} It is defined as the number of matched bits between the secret \( s \) and the decoded secret \( \texttt{Dec}(x_a) \) for an attacked image \( x_a \):
\[
\# \text{ Matched Bits} = \sum_{i=1}^{l} \mathbb{I} \left( s[i] = \texttt{Dec}(x_a)[i] \right)
\]

where \( s[i] \) is the \( i \)-th bit of the secret $s$, \( \texttt{Dec}(x_a)[i] \) is the \( i \)-th bit of the decoded secret, and \( \mathbb{I}(a = b) \) is an indicator function that equals 1 if \( a = b \) and 0 otherwise. 

\vspace{1mm}
\noindent{\bf 4. Perceptibility:} For quality, we additionally report {LPIPS}~\citep{zhang2018unreasonable} for quantitative perceptual similarity. We also report $l_\infty$ and SSIM in Appendix \ref{additional_metrics}.

\vspace{1mm}
\noindent{\bf 5. Runtime:} We compare the running time (attack time only, not including time taken for training surrogate models) across different methods. 

For metrics over different model checkpoints, we plot the mean and the range from the minimum to the maximum.

\subsubsection{Other Details}

\noindent{\bf \# Surrogate Models:} We use $k \in \{1, 2, 5, 10, 20, 30, 40, 50\}$. We do not consider larger values as the runtime of the attack of~\citet{hu2024transfer} increases rapidly with more surrogate models.

\vspace{1mm}
\noindent{\bf Evaluation Size:} The size of the dataset for evaluation is $N=1,000$ images. 

\vspace{1mm}
\noindent{\bf Baseline Implementations:} For DiffPure, we set the same diffusion steps $t \in \{0.1,0.2,0.3\}$, but add an additional $t=0.05$ to consider the case for even smaller perturbation. For the transfer attack~\citep{hu2024transfer}, we use the implementation shared by the authors and keep parameters unchanged. Image pixels are in $[-1,1]$ range. We ensure adding perturbations never gets pixel values outbound by clamping them accordingly before adding them to target images.

\vspace{1mm}
\noindent{\bf Attack Budget:} We set the $\ell_\infty$ budget to be $0.25$.

\section{Results}
\label{sec:results}

\subsection{Naive Transfer: No Optimization}
\label{sec:naive}

We perform a naive transfer attack by using OFT on different target models with $k=1$. This reflects the case where configurations are not aligned and the attacker has limited computation capability. Only the control group, i.e., when the target model is the same as the surrogate model concludes with a successful attack. The results are presented in Appendix~\ref{app:naive}. This further motivates us to design experiments relaxing one single assumption at a time.

\subsection{Transfer with Unaligned Configurations}
\label{sec:exp1}

\begin{figure}[h]
\begin{subfigure}{0.48\columnwidth}
\includegraphics[width=\linewidth]{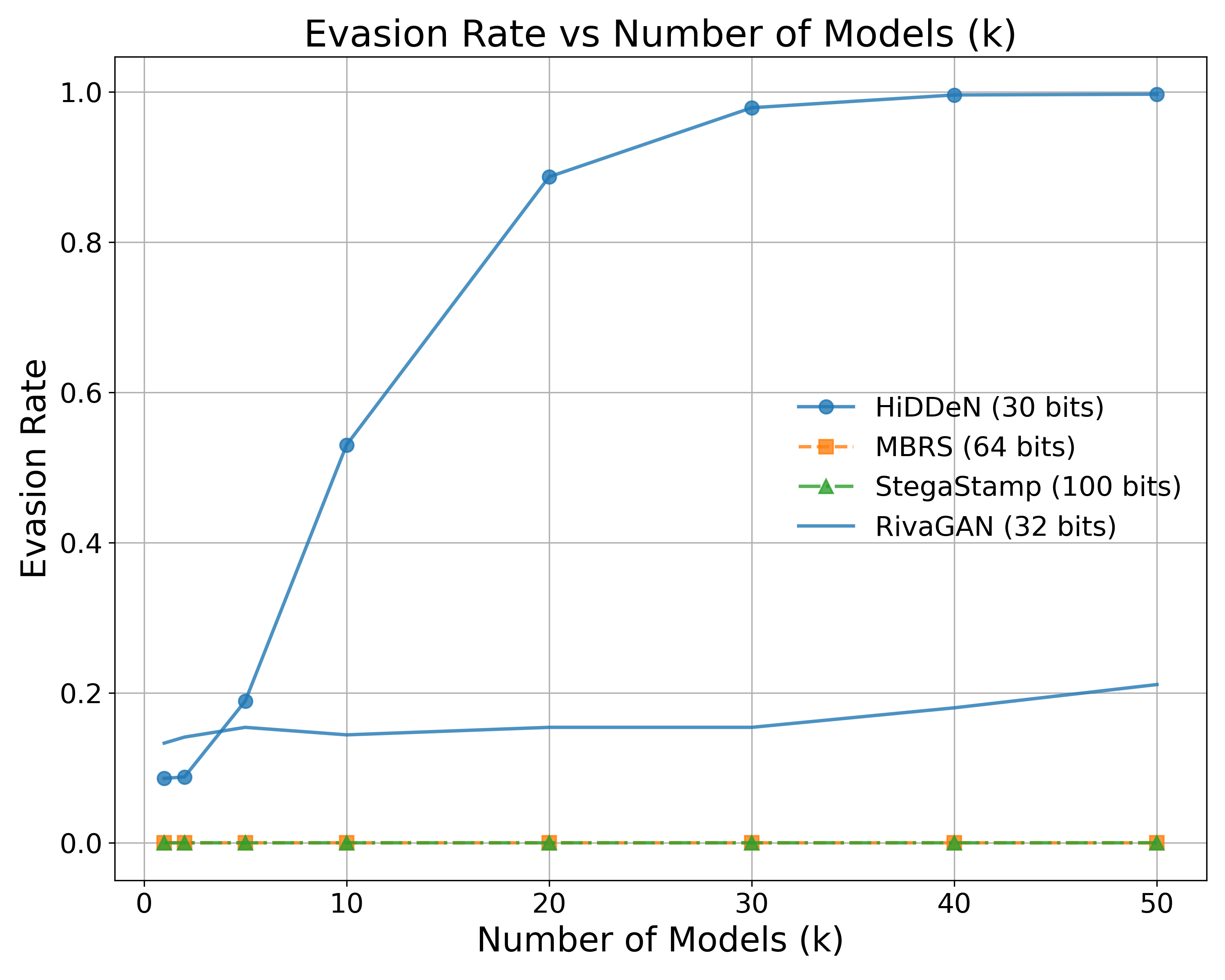}
\caption{{\bf Evasion rate}}
\label{unaligned}
\end{subfigure}
\hfill
\begin{subfigure}{0.48\columnwidth}
\includegraphics[width=\linewidth]{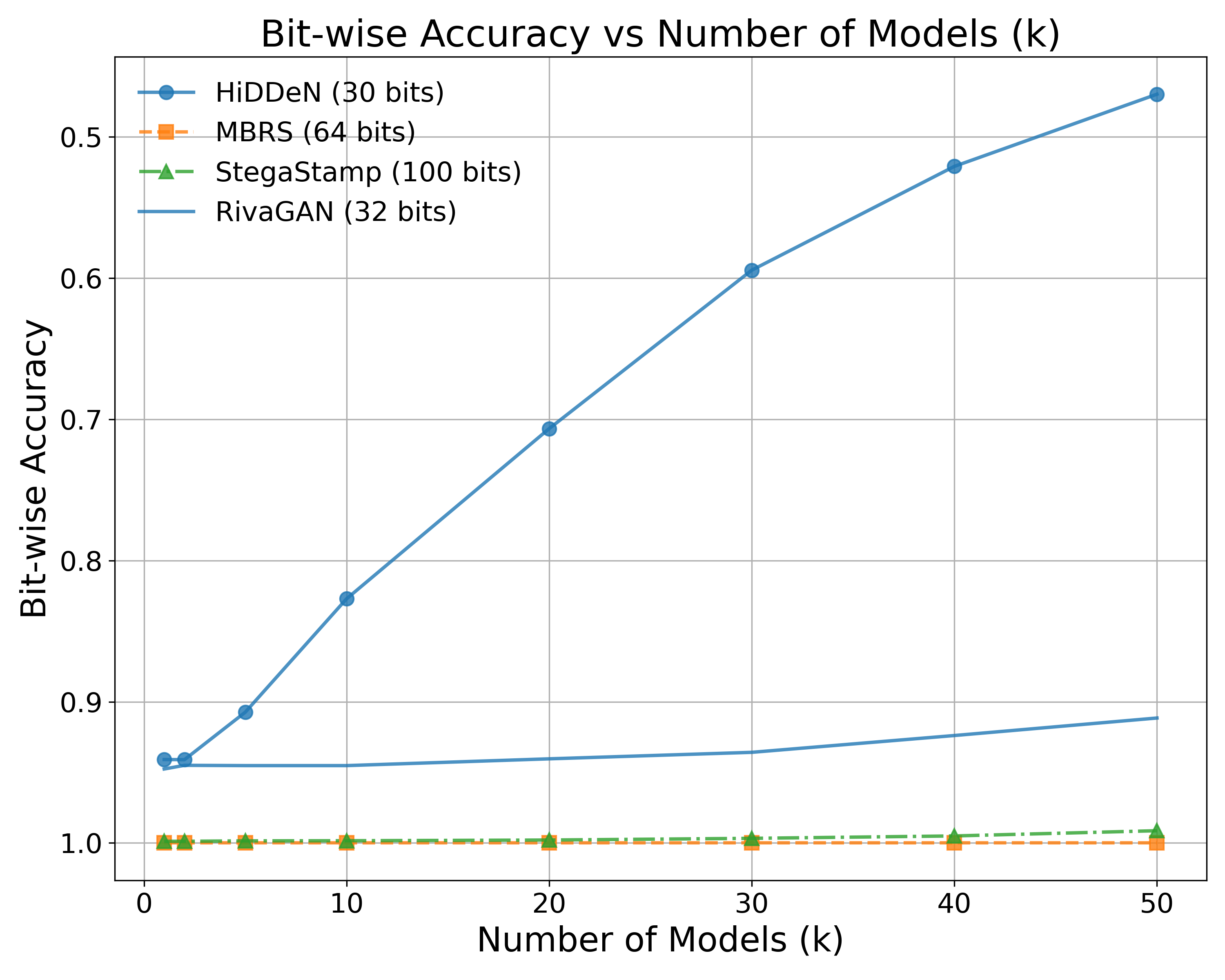}
\caption{{\bf \texttt{BA}}}
\label{bitwise}
\end{subfigure}
\caption{{\bf Evaluation of~\citet{hu2024transfer}'s attack to different watermarking methods.} The only successful attack is when the target model's method matches the surrogate models' method (\texttt{HiDDeN}).}
\label{fig:combined}
\end{figure}

\begin{table*}[t]
\centering
\scriptsize
\caption{{\bf Comparison of Performance Metrics:} The left portion of the table corresponds to information for Fig.~\ref{Image Samples}, wheras the right corresponds to all $N=1,000$ samples.}
\begin{tabular}{l@{\hspace{2em}}cccc@{\hspace{2em}}cccc}
\cmidrule(l{0pt}r{2em}){1-1} \cmidrule(l{0pt}r{2em}){2-5} \cmidrule(l{0pt}){6-9}
\textbf{Method} & \textbf{\# Matched Bits} & \textbf{Detected?} & \textbf{LPIPS} & \textbf{$l_\infty$} & \textbf{Overall Bit-wise Acc.} & \textbf{Evasion Rate} & \textbf{Overall LPIPS} & \textbf{Overall $l_\infty$} \\
\cmidrule(l{0pt}r{2em}){1-1} \cmidrule(l{0pt}r{2em}){2-5} \cmidrule(l{0pt}){6-9}
\cmidrule(l{0pt}r{2em}){1-1} \cmidrule(l{0pt}r{2em}){2-5} \cmidrule(l{0pt}){6-9}
Unattacked & 29 & \ding{51} & - & - & - & 0.005 & - & -\\
\citet{hu2024transfer} $k=50$ & 12 & \ding{55} &0.0043 &0.2500 & 0.4702&0.998 &0.0177 &0.1951 \\
DiffPure ($t=0.05$) \citep{saberi2023robustness} & 26 & \ding{51} & 0.0506 & 0.4961 & 0.8078 & 0.690 & 0.0797 & 0.3410\\
DiffPure ($t=0.1$) \citep{saberi2023robustness} & 22 & \ding{55} & 0.0591 & 0.5662 & 0.6802 & 0.974 & 0.1242 & 0.4482 \\
DiffPure ($t=0.2$) \citep{saberi2023robustness} & 16 & \ding{55} & 0.0817 & 0.6707 & 0.5685 & 1.000 & 0.1623 & 0.6155 \\
DiffPure ($t=0.3$) \citep{saberi2023robustness} & 16 & \ding{55} & 0.1087 & 0.8732 & 0.5299 & 1.000 & 0.1921 & 0.7966 \\
OFT (Unnorm. $k=1$) & 17 & \ding{55} & 0.0763 & 0.4954 & 0.4907 & 1.000 & 0.0781 & 0.4682\\
OFT (Norm. $k=1$) & 21 & \ding{55} & 0.0709 & 0.2500 & 0.4974 & 1.000 & 0.0767 & 0.2488\\
\cmidrule(l{0pt}r{2em}){1-1} \cmidrule(l{0pt}r{2em}){2-5} \cmidrule(l{0pt}){6-9}
\end{tabular}
\label{Metrics}
\end{table*}

We compare the evasion rate when evaluating the transfer attack~\citep{hu2024transfer} on different target models: \texttt{HiDDeN}, \texttt{StegaStamp}, \texttt{RivaGAN}, \texttt{MBRS}. As the surrogate models are \texttt{HiDDeN} models, \texttt{HiDDeN} is the contrast group. Results are shown in Fig.~\ref{unaligned}. Observe that the evasion rate and \texttt{BA} (in Fig.~\ref{bitwise}) clearly show that the attack of~\citet{hu2024transfer} {\em cannot work on a different architecture/method compared with the surrogate models}. It is worth noting that~\citet{hu2024transfer} has a similar experiment transferring to different watermark methods showing that it is still effective (refer Fig.6 in~\citet{hu2024transfer}). However, in that experiment, both \texttt{HiDDeN} and \texttt{StegaStamp} are included in the surrogate models, while the targets are (smoothed \footnote{\citet{jiang2024certifiably}. In smoothing, images are perturbed by Gaussian noise for a batch of perturbed version before sending to the detector, then the detection results depends on the median \texttt{BA} of these perturbed images} and vanilla) \texttt{HiDDeN}, \texttt{StegaStamp} and \texttt{Stable Signature\citep{fernandez2023stable}}, whose watermark is learned from fine-tuning on images watermarked by \texttt{HiDDeN}. This suggests that even powerful optimization-based attacks fail when the surrogates are not aligned.

\subsection{Efficacy of OFT}
\label{sec:exp2}

\begin{figure}[h]
\begin{center}
\includegraphics[width=\linewidth]{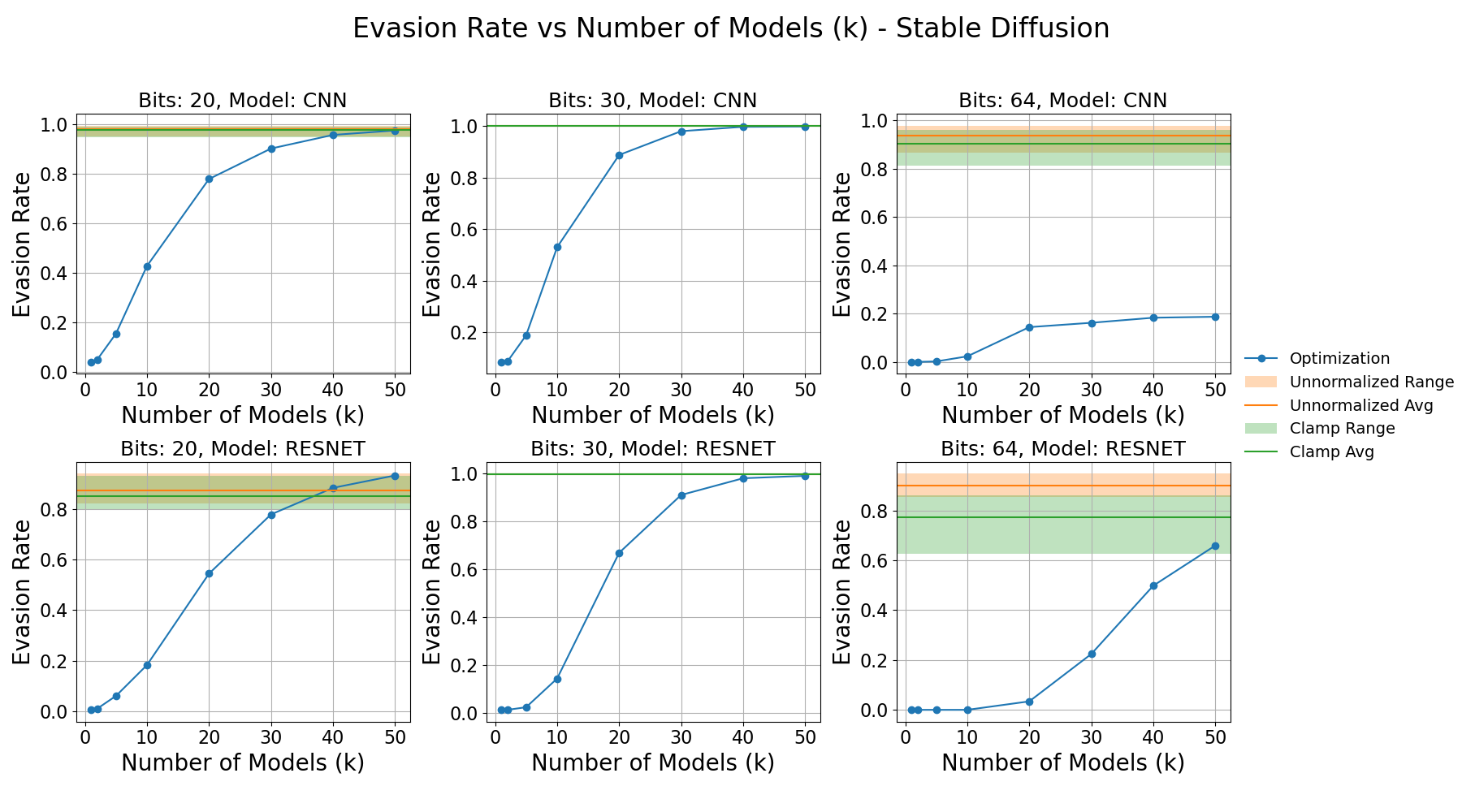} 
\end{center}
\caption{{\bf Evasion rate comparing \citet{hu2024transfer} and OFT ($k=1$):} Observe that OFT is superior most often!}
\label{fig:multi_evasion}
\end{figure}

We now evaluate OFT w.r.t the attack proposed by~\citet{hu2024transfer}. For OFT, we only use $k=1$ surrogate model, but vary the training of the single model (resulting in a ``region'' of evasion rates and bit-wise accuracies). All presented results are when the models are trained with the DiffusionDB dataset~\citep{wang2022diffusiondb}; results are consistent with the MidJourney dataset~\citep{iulia_turc_gaurav_nemade_2022} too (c.f., Appendix~\ref{appendix:mjresults}). The surrogate and target models are for \texttt{HiDDeN}, and the evasion rates are presented in Fig.~\ref{fig:multi_evasion}. The top row corresponds to the CNN architecture, and the bottom corresponds to ResNet. From Fig.~\ref{fig:multi_evasion}, we can clearly see that even if we compare the best configuration ($k=50$) for~\citet{hu2024transfer}'s attack with OFT normalized (green region), it does not have an advantage except in the $\ell=20$, ResNet architecture scenario; even there, the advantage is limited. 

\begin{figure}[h]
\begin{center}
\includegraphics[width=\linewidth]{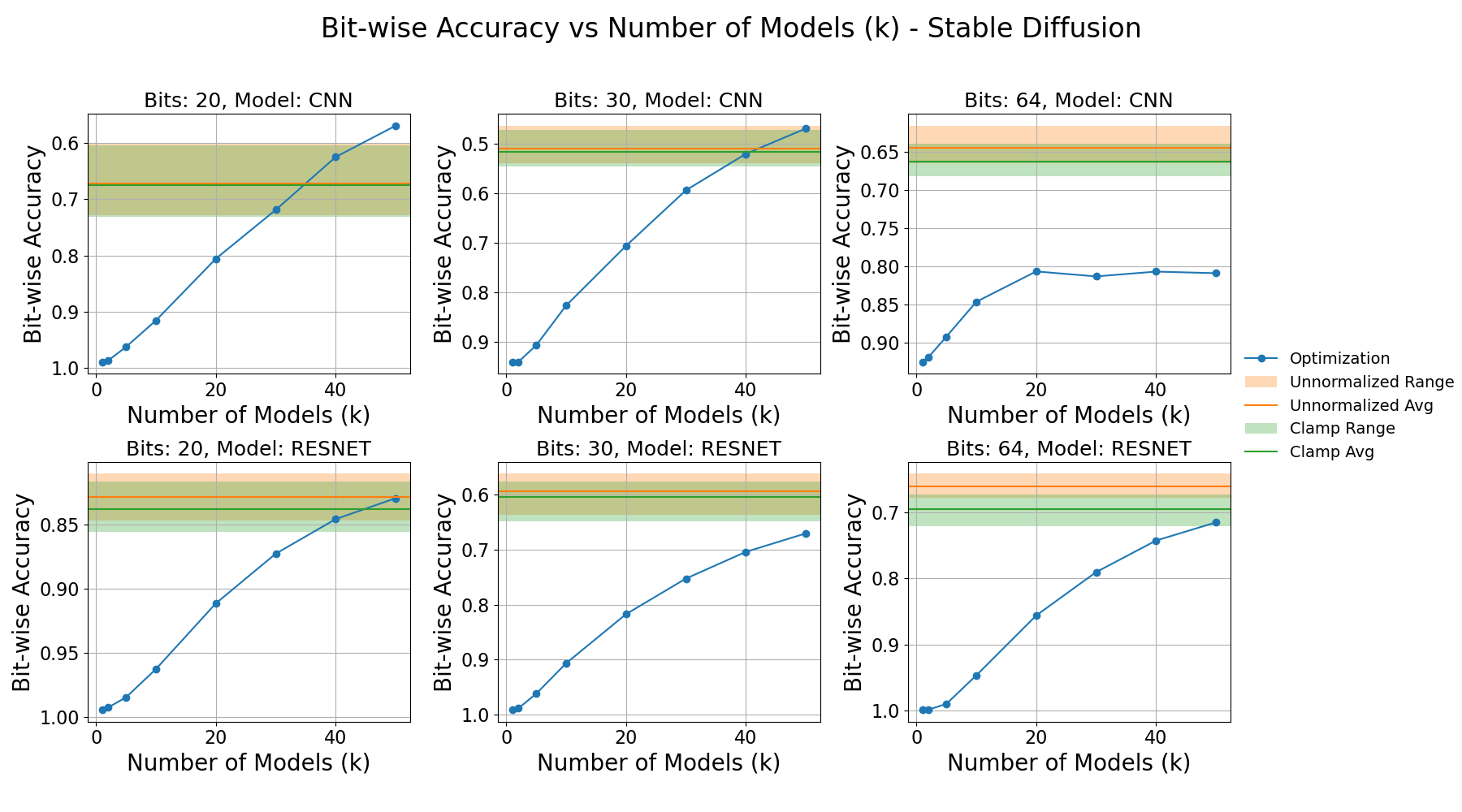} 
\end{center}
\caption{{\bf Bit-wise accuracy comparing \citet{hu2024transfer} and OFT ($k=1$):} Observe that OFT is superior most often!}
\label{fig:multi_ba}
\end{figure}

Bit-wise accuracy, reported in Fig.~\ref{fig:multi_ba} (CNN top row, ResNet bottom row), also shows a similar trend. Note that while~\citet{hu2024transfer}'s method has some advantage for $50$ surrogate models, it does not reflect in the evasion rate.

Besides, we can also see that \citet{hu2024transfer}'s attack is much worse than ours for $\ell=64$ bits with the CNN architecture. It is likely due to the change in hyperparameters; such a change significantly impacts~\citet{hu2024transfer}'s attack but not ours. Yet, the width of colored areas in Figure \ref{fig:multi_evasion} and~\ref{fig:multi_ba} shows that the variance over different checkpoints is not ignorable. 

Note that besides the green area, we also plot the unnormalized version of our attack, which can go beyond $0.25$ $\ell_\infty$ budget. The plots show that normalization has a limited effect weakening our attack.

\subsection{Limited Number of Surrogate Models}

If we consider scenarios where $k < 5$ in Figs.~\ref{fig:multi_evasion} and~\ref{fig:multi_ba} for~\citet{hu2024transfer}'s attack, it is clear they are not successful (with close to $0$ evasion rate and close to $1$ bit-wise accuracy). Consequently, multiple surrogate models are essential for the success of~\citet{hu2024transfer}'s attack. In contrast, {\em our attack only needs one surrogate model.}

\subsection{Microbenchmarks}

\subsubsection{Runtime Comparison}

\begin{table}[h]
\centering
\caption{{\bf Runtime Comparison:} OFT leads the pack by a mile!}
\begin{tabular}{lc}
\toprule
\textbf{Methods} & \textbf{Time (s)} \\
\midrule
\midrule
\citet{hu2024transfer}, $k=50$ & 2493.0594\\
Diffpure~\cite{{saberi2023robustness}} ($t=0.05$) & 69.7701 \\
Diffpure~\cite{{saberi2023robustness}} ($t=0.1$) & 138.3874 \\
Diffpure~\cite{{saberi2023robustness}} ($t=0.2$) & 274.6078 \\
Diffpure~\cite{{saberi2023robustness}} ($t=0.3$) & 411.7274 \\
OFT (unnormalized, $k=1$) & \textbf{0.3853} \\
OFT (normalized, $k=1$) & {0.3987} \\
\bottomrule
\end{tabular}
\label{RT}
\end{table}

While expected, from Table~\ref{RT}, it is clear that OFT ($k=1$) is the fastest: attacking $1,000$ images takes $<1$ second. The most effective version ($k=50$) of~\citet{hu2024transfer} is the slowest (by nearly 3 orders of magnitude), while DiffPure~\citep{saberi2023robustness} is in the middle, and the time scales roughly proportional to the diffusion step ($t$) hyperparameter.

\subsubsection{Image Quality Comparison}
\label{sec:quality}

Visuals of OFT's outputs is shown in Fig.\ref{Image Samples}. For quantitative analysis, we report the quality metrics for the example in Fig.~\ref{Image Samples} along with detection results, and the corresponding metrics over all tested samples in Table~\ref{Metrics}. Under the setting where the surrogate and the target model's configurations are mostly aligned, our method can better preserve the image quality compared with DiffPure~\citep{saberi2023robustness} by having a lower LPIPS and $l_\infty$, though it is worse than~\citet{hu2024transfer}. It does so while having comparable attack efficacy to~\citet{hu2024transfer}, while being substantially less expensive.

\section{Discussion}
\label{sec:discussion}

\subsection{Limitations}

One limitation of the simple attack we propose is that it is heuristic and there is no guarantee for evasion. While there are bounds on \texttt{BA} for attacked images from~\citet{hu2024transfer}, its estimation is based on estimating the unperturbed, positive transferring, and negative transferring similarity (defined in Appendix F of their draft \footnote{\url{https://arxiv.org/abs/2403.15365}}) between surrogate models and the victim model. However, such estimation is not available in the no-box setting. Also, note that as perturbation added by OFT can be seen as a solution $\delta$ to their formulated problem (equation (3) in \citet{hu2024transfer}\footnote{\url{https://arxiv.org/abs/2403.15365}}, as the \texttt{BA} constraint should hold for some reasonable sensitivity and given $r$), their derivation can be applied for similar bounds for OFT.

Another limitation is that the variance over different checkpoints is not small. Such a variance could be larger than that for~\citet{hu2024transfer}, whose variance over various sets of surrogate models is not presented due to the high computational cost of their attack.

\subsection{Other Watermarking Methods}

The watermark methods we investigate are learning-based and general-purpose. Non-learning-based watermarks like \texttt{DWT-DCT}~\citep{al2007combined} are less robust in general. There are also generator-specific methods like those proposed by~\citet{fernandez2023stable} that incorporate watermarking in image generation models so that generated images are automatically watermarked. These methods' models cannot be directly included in surrogate models as they do not have an independent encoder to watermark images.

\section{Conclusions}
\label{sec:conclusions}
Our study presents that transfer-based attack is challenging under the no-box setting. We demonstrate that the current method depends on impractical assumptions heavily and find that optimization has a limited advantage. This reveals the fundamental challenge for transfer-based attack on image watermarking in no-box setting: limited knowledge on the similarity between the surrogate model(s) and the victim model. We aim to inspire further research on more practical and careful evaluation to no-box attacks on image watermarking. 

\newpage

{
    \small
    \bibliographystyle{ieeenat_fullname}
    \bibliography{main}
}


\appendix
\onecolumn
\section*{Appendix}
\section{Details: Watermarking Methods}
\label{sec:appendix_wm}

Different watermarking methods architectures can be very different. Here we list their architecture details for their implementations. The overall encoder is defined as follows:
$\texttt{Enc}(s,x)=f(h(e(x),g(s)),x)$

\subsection{\texttt{HiDDeN}~\citep{zhu2018hidden}}
\label{ss:hidden_arch}

\texttt{HiDDeN} comprises four main components: an encoder, a parameterless noise layer, a decoder, and an adversarial discriminator. The encoder, decoder, and discriminator have trainable parameters. The encoder receives a cover image of shape $3 \times w \times h$ and a binary secret message of length $\ell$. It produces an encoded image $x'$ of the same shape as the cover image. The noise layer then receives the encoded image as input and adds distortions to produce a noised image. The decoder aims to recover a message from the noised image. At the same time, given either a cover image or an encoded image, the adversarial discriminator predicts the probability that the image is an encoded image. Feedback from this discriminator is used to improve the perceptual quality of the encoded image (making it more similar to the cover image).

\citet{zhu2018hidden} also imposes a message distortion loss, using the squared Euclidean distance between the original and decoded messages, normalized by the message length. Both the encoder and decoder are jointly trained, whilst simultaneously training the discriminator to minimize its classification loss over the same distribution of input messages and images. 10,000 and 1,000 images from the COCO dataset~\citep{lin2014microsoft} are used as the training and validation set to train the official checkpoint. In our experiments, the surrogate models are trained on DALLE-2~\citep{dalle_gallery_2023}, and the target models are trained on DiffusionDB~\citep{wang2022diffusiondb} or MidJouruney~\citep{iulia_turc_gaurav_nemade_2022}, with the same number of images for the training and validation dataset. 

Using the notation as described in \S~\ref{sec:general}, the encoder of \texttt{HiDDeN} can be described as below\footnote{We use the shorthand \texttt{Conv-BN-ReLU} to denote $\texttt{Conv} \circ \texttt{BN} \circ \texttt{ReLU}$}. Here $e_{ResNet}$ stands for the architecture used by~\citet{hu2024transfer}: an alternative structure they use in experiments compared with CNN, and the only difference is 7 instead of 4 \texttt{Conv-BN-ReLU} layers in $e$:
\begin{itemize}
    \item Image dimension: $w=h=128$ 
    \item $\ell=30$ (by default, can also be $20$, $64$)
    \item $x \in [0,1]^{3 \times w \times h}$
    \item $e_{CNN}$: 4 \texttt{Conv-BN-ReLU} layers, $x_e=(\texttt{Conv-BN-ReLU})^4(x)$; $x_e \in \mathbb{R}^{64 \times w \times h}$

    \item $e_{ResNet}$ 7 \texttt{Conv-BN-ReLU} layers, $x_e=(\texttt{Conv-BN-ReLU})^7(x)$; $x_e \in \mathbb{R}^{64 \times w \times h}$

    \item $g$: scaling up via self-concatenation i.e., $\{0,1\}^{\ell} \rightarrow \{0,1\}^{\ell \times w \times h}$; $s_g \in \{0,1\}^{\ell \times w \times h}$

    \item $h$: concatenation, $x_h=x_e \| s_g$ s.t. $x_h \in \mathbb{R}^{(64 + \ell) \times w \times h}$

    \item $f$: \texttt{Conv-BN-ReLU} and \texttt{Conv2D} i.e., $f(x_h,x)= \texttt{Conv2D} \circ \texttt{Conv-BN-ReLU}(x_h\|x))$; $x_f \in [0,1]^{3 \times w \times h}$ 
\end{itemize}

\medskip
The decoder architecture is as follows:
\begin{itemize}
\item \texttt{Dec$_{CNN}$}: 8 \texttt{Conv-BN-ReLU} layers followed by \texttt{AdaptiveAvgPool2d} and a fully-connected layer, $s'=\texttt{Dec}(x) = \texttt{Linear}\circ \texttt{AdaptiveAvgPool2d} \circ (\texttt{Conv-BN-ReLU})^{8}$; $s' \in \{0,1\}^{\ell}$
\item \texttt{Dec$_{ResNet}$}: \texttt{ResNet-18}, $s'=\texttt{Dec}(x)=\texttt{ResNet-18}(x)$; $s' \in \{0,1\}^{\ell}$
\end{itemize}

\subsection{\texttt{MBRS}~\citep{jia2021mbrs}}
\label{ss:mbrs}

\texttt{MBRS}'s framework is the same as \texttt{HiDDeN}. The main difference or novelty is the JPEG-related augmentation in the noise layer. Besides, the secret is encoded by \texttt{Conv-BN-ReLU} and \texttt{SENet}~\citep{hu2018squeeze}, rather than simply scaled up as in \texttt{HiDDeN}. A watermark strength factor $c$ is involved in adjusting the distortion. For the official checkpoint, 10,000 images from ImageNet~\citep{deng2009imagenet} and 5,000 images from COCO~\citep{lin2014microsoft} are used as training and validation set, respectively. For the partially aligned version, 10,000 and 1,000 images from DALLE-2 dataset~\citep{dalle_gallery_2023} are used for the training and validation dataset to align to \texttt{HiDDeN} surrogate models.

Using the notation as described in \S~\ref{sec:general}, the encoder of \texttt{MBRS} can be described as below:
\begin{itemize}
    \item Image dimension: $w=h=256$ for the official checkpoint and $w=h=128$ for the partially aligned version
    \item $\ell=256$ for the official checkpoint and $\ell=64$ for the partially aligned version
    \item $x \in [0,1]^{3 \times w \times h}$
    \item $e$: \texttt{Conv-BN-ReLU} and \texttt{SENet~\citep{hu2018squeeze}}, $x_e=\texttt{SENet} \circ \texttt{Conv-BN-ReLU}(x)$;
    $x_e \in \mathbb{R}^{64 \times w \times h}$ 
    
    \item $g$: \texttt{Conv-BN-ReLU}, \texttt{ExpandNet} (also \texttt{Conv-BN-ReLU} to scale up) and 2 \texttt{SENet}, 
    $g(s)=(\texttt{SENet})^2\circ \texttt{ExpandNet} \circ \texttt{Conv-BN-ReLU}(s)$;
    $s_g \in \mathbb{R}^{\ell \times w \times h}$
    
    \item $h$: \texttt{Conv-BN-ReLU}, $x_h=\texttt{Conv-BN-ReLU}(x_e \| s_g)$;
    $x_h \in \mathbb{R}^{64 \times w \times h}$
    
    \item $f$: \texttt{Conv2D} with concatenation involving strength factor $c$, $x'=x+c(\texttt{Conv2D}(x_h\|x)-x)$
\end{itemize}

\medskip
The decoder architecture is as follows:
\begin{itemize}
    \item \texttt{Dec}: \texttt{Conv-BN-ReLU} and \texttt{SENet\_decoder} blocks, $s' = \texttt{Dec}(x) = (\texttt{Conv-BN-ReLU})^2 \circ (\texttt{SENet\_decoder})^5 \circ \texttt{Conv-BN-ReLU}$; $s' \in \{0,1\}^{\ell}$
\end{itemize}

\subsection{\texttt{RivaGAN}~\citep{zhang2019robust}}

\texttt{RivaGAN} is an attention-based watermarking method for images and videos. For ease of comparison, we focus on the image case and adapt our notations. Similar to \texttt{HiDDeN}, it has five components: an encoder, noise layers, a decoder, a critic, and an adversary. The first four components have trainable parameters. Attention is the important module used in both the encoder and decoder. It receives a cover image of shape $1 \times w \times h \times 3$ and uses two convolutional blocks to create an attention mask of shape $1 \times w \times h \times \ell$, which is then used by the encoder and decoder modules to determine which bits to pay attention to at each pixel. The encoder module uses the attention mask to compute a compacted form of the data tensor and concatenates it to the image before applying additional convolutional blocks to generate the encoding residual. Then the encoding residual is added to the input (cover) image for the final watermarked image. Then the noise layer receives it and distorts the watermarked image. The decoder module extracts the data of shape $1 \times w \times h \times \ell$ but then weights the prediction using the attention mask from the attention module before averaging to try and recover the original data. At the same time, given either a cover image or an encoded image, the adversary network attempts to imitate an attacker trying to remove the watermark, while the critic network is responsible for taking a sequence of images and detecting the presence of a watermark. Feedback from this adversary and critic network is used to improve the robustness of the watermark and perceptual quality of the encoded image (making it more similar to the cover image), respectively. 

\citet{zhang2019robust} also imposes a message distortion loss, using the cross entropy between the original and decoded messages, normalized by the message length. Both the encoder and decoder are jointly trained, whilst simultaneously training the critic and the adversary to minimize their loss over the same distribution of input messages and images and maximize robustness. \texttt{Hollywood 2}~\citep{marszalek2009actions} is used as the training and validation set, including 3669 video clips and approximately 20.1 hours of video in total.

Using the notation as described in \S~\ref{sec:general}, the encoder of \texttt{RivaGAN} can be described as below\footnote{We use the shorthand \texttt{Conv3d-Tanh-BatchNorm3d} to denote $\texttt{Conv3d} \circ \texttt{Tanh} \circ \texttt{BatchNorm3d}$}:

\begin{itemize}
    \item $\ell=32$
    \item $w=h=256$ for image dimension
    \item $x \in [0,1]^{1 \times w \times h \times 3}$
    \item $c=0.01$ for watermark strength
    \item $e$: attention mask by 2 \texttt{Conv3d-Tanh-BatchNorm3d} and \texttt{Softmax}, $x_e=\texttt{Softmax} \circ (\texttt{Conv3d-Tanh-BatchNorm3d})^2(x)$; $x_e \in \mathbb{R}^{1 \times w \times h \times \ell}$
    \item $g$: identity, $s_g=s$; $s \in \{0,1\}^{\ell}$
    \item $h$: dot product on the last dimension with broadcast, $x_h=x_e \cdot s_g =x_e \cdot s$; $x_h \in \mathbb{R}^{1\times w \times h \times 1}$
    \item $f$: 2 \texttt{Conv3d-Tanh-BatchNorm3d} involving strength factor $0.01$, $x'=x+c \times \texttt{Tanh} \circ (\texttt{Conv3d-Tanh-BatchNorm3d})^2 (x_h \| x)$
\end{itemize}

\medskip
The decoder architecture is as follows:
\begin{itemize}
    \item \texttt{Dec}: average on the dot product of convolutional features and attention obtained from the same attention mask as $e$, $s' = \texttt{Dec}(x) = \texttt{Average} \circ ((\texttt{Conv3d-Tanh-BatchNorm3d})^2(x) \cdot e(x))$; $s' \in \{0,1\}^{\ell}$
\end{itemize}

\subsection{\texttt{StegaStamp}~\citep{tancik2020stegastamp}}
\label{ss:ss_arch}
\texttt{StegaStamp} comprises five main components: an encoder, a parameterless noise layer, a detector, a decoder, and a critic network. The encoder, detector, decoder, and critic have trainable parameters. The encoder works similarly as in \texttt{HiDDeN}. The noise layer receives both the cover image and the encoded image as input. It adds distortions, simulating the corruptions caused by printing, reimaging, and detecting the \texttt{StegaStamp} with a set of differentiable image augmentations. The image detector and decoder replace the decoder in \texttt{HiDDeN} for watermark detection, where the image detector is newly added. For the image detector, they fine-tune an off-the-shelf semantic segmentation network BiSeNet~\citep{yu2018bisenet} to segment areas of the image that are believed to contain \texttt{StegaStamps}, which is trained on DIV2K~\citep{agustsson2017ntire}. The critic network works similarly to the decoder and the adversarial discriminator in \texttt{HiDDeN}. \citet{tancik2020stegastamp} also impose a message distortion loss, using cross entropy between the original and decoded messages, normalized by the message length. The encoder and decoder (and the image detector) are jointly trained (fine-tuned), whilst simultaneously training the discriminator to minimize its classification loss over the same distribution of input messages and images. During training, they use images from the MIRFLICKR dataset~\citep{huiskes2008mir} (rescaled to $400 \times 400$ resolution) combined with randomly sampled binary messages.

Using the notation as described in \S~\ref{sec:general}, the encoder of \texttt{StegaStamp} can be described as below:\footnote{We use the shorthand \texttt{Conv2DReLU} to denote $\texttt{Conv2D} \circ \texttt{ReLU}$}

\begin{itemize}
    \item $\ell=100$
    \item $w=h=400$ for image dimension
    \item $x \in [0,1]^{3 \times w \times h}$
    \item $e$: normalize to $[-0.5,0.5]$ from $[0,1]$, $x_e=x-0.5$
    \item $g$: fully connected layer with Kaiming Initialization~\citet{he2015delving}, reshape and up-sample, $s_g=\texttt{Upsample}\circ \texttt{Reshape} \circ \texttt {FC} (s-0.5)$; $s\in [0,\infty)^{3*w*h}$
    \item $h$: (U-Net\footnote{O. Ronneberger, P. Fischer, and T. Brox, "U-net: Convolutional networks for biomedical image segmentation," in *Medical Image Computing and Computer-Assisted Intervention – MICCAI 2015: 18th International Conference, Munich, Germany, October 5-9, 2015, Proceedings, Part III*, vol. 9351, Springer International Publishing, 2015, pp. 234–241.
}
    
    style architecture) $10$ layers of \texttt{Conv2D} including $4$ up-sampling in-between with shortcuts, $x_h=\texttt{Conv2DReLU layers with up-sampling and shortcuts}(x_e\|s_g)$ ; $x_h \in \mathbb{R}^{3 \times w \times h}$

    \item $f$: add (residual) $x_h$ to the original image, $x'=x+x_h$

\end{itemize}

\medskip
The decoder architecture is as follows:
\begin{itemize}
    \item \texttt{Dec}: normalized $x$ goes thorough \texttt{SpatialTransformerNetwork}\footnote{M. Jaderberg, K. Simonyan, and A. Zisserman, "Spatial transformer networks," in *Advances in Neural Information Processing Systems 28 (NeurIPS)*, 2015.}
    , \texttt{Conv2DReLU} layers, \texttt{Linear} layers with \texttt{ReLU} activation and \texttt{Sigmoid}; $s' = \texttt{Dec}(x) = \texttt{Sigmoid} \circ \texttt{Linear} \circ \texttt{ReLU} \circ \texttt{Linear} \circ \texttt{Flatten} \circ (\texttt{Conv2DReLU})^7 \circ \texttt{SpatialTransformerNetwork}(x-0.5)$, where $s' \in \{0,1\}^{\ell}$
\end{itemize}

\section{Attacks on Image Watermarks}
\label{app:attack_models}

Here we list more details on different attack methods on image watermarks, including representative no-box, black-box, and no-box attacks that are less effective and not discussed in \S~\ref{subsec:no_box_examples}.

Recall that from white-box to no-box, the attacker has progressively lesser access to the detector, making the attacks progressively harder, but more and more practical. While white-box attacks test the worst-case guarantees of the system, these are not realistic in many situations as those deploying watermarking hardly allow white-box access to the detector. Black-box is more practical, yet requires the detector's API or other black-box assumptions, while no-box has the least access requirements.

\subsection{White-box}

In the white-box setting, the attacker has full unrestricted access to the watermark detector. The white-box access significantly reduces the difficulty for the attacker. For example, WEvade-W-II \citep{jiang2023evading} can evade two-tail detection using optimization. 

\subsection{Black-box}

In the black-box setting, the attacker has limited access to the detector, meaning the attacker can only query the detector API to get binary (watermarked / unwatermarked) results. Alternative formulations involve the attacker having access to a set of images for both the watermarked and unwatermarked versions of the model. Several methods have been proposed in the black-box setting. WEvade-B-S~\citep{jiang2023evading}, uses a surrogate encoder to convert to the white-box setting. HopSkipJump~\citep{chen2020hopskipjumpattack} can be used to evade image watermarking by random initialization and then walking while guaranteeing evasion until the query budget is used up. WEvade-B-Q~\citep{jiang2023evading} upgrades HopSkipJump by replacing its random initialization with JPEG initialization and introducing early stopping.  Also,~\citet{lukas2023leveraging} leverages optimization for an adaptive attack.~\citet{an2024box} leverages images and their watermarked version pairs for black-box image watermarks, which is black-box as watermarked-unwatermarked image pairs are assumed to be available.

\subsection{Additional No-Box Attacks}

\noindent{\bf Non-learning Distortions:} Single and combined non-learning distortions (like cropping, JPEG noise) are also evaluated in~\citet{an2024box} as attack methods. However,~\citet{an2024box} shows that the learning-based watermarks they evaluate, i.e., \texttt{Tree-Ring}~\citep{wen2023tree}
. \texttt{Stable Signature}~\citep{fernandez2023stable}, \texttt{StegaStamp}, and \texttt{MBRS}, are generally robust against this type of attacks, though \texttt{MBRS} is not robust to some distortions.

\smallskip
\noindent{\bf Embedding Attacks:} Embedding attacks in~\citet{an2024box} crafts adversarial image $x_{a}$ to diverge its embedding from the original image $x_{wm}$ but within $\ell_{\infty}$ perturbation budget in the image space. This is successful on \texttt{Tree-Ring} but not on \texttt{Stable Signature}~\citep{fernandez2023stable}, \texttt{StegaStamp}, and \texttt{MBRS}.

\smallskip
\noindent{\bf StegaAnalysis:}~\citet{yang2024steganalysis} assumes the availability of unwatermarked images which can be paired or unpaired with the target watermarked images. For the unpaired variant, such unwatermarked images can be from another dataset, so we consider this variant no-box. This method tries to extract watermark patterns from watermarked and unwatermarked images to subtract the extracted patterns for removal, but its success is mainly limited to content-agnostic watermarks. Content-agnostic methods use fixed, predefined watermark patterns independent of or weakly dependent on image content, which are usually more high-perturbation and perceptible. 
\smallskip
\noindent{\bf Surrogate Detector Attacks:} \texttt{AdvCls-Real\&WM} from~\citet{an2024box} is the only attack considered no-box within their surrogate detector attacks as it uses watermarked images and unwatermarked images to train a surrogate detector; unwatermarked images are sampled ImageNet~\citep{deng2009imagenet} dataset rather than the victim model. Then attacks are conducted on the surrogate model. However,~\citet{an2024box} finds it is only successful on \texttt{Tree-Ring} but not \texttt{Stable Signature}, \texttt{StegaStamp}, and \texttt{MBRS}.

\smallskip
\noindent{\bf UnMarker:} Another concurrent work~\citep{kassis2024unmarker} uses spectral optimization and has successful evasion. However, cropping is one necessary component for some target watermarking methods, and quality degradation is visible, which can be undesirable in some scenarios and fails norm constraints, and the computation cost is not cheap. 

\section{Additional Results}

\subsection{Additional Metrics}
\label{additional_metrics}

We additionally report the following metrics for \S\ref{sec:exp2}
\begin{enumerate}
    \item $\ell_\infty$, defined as:
$\|x_a-x_{wm}\|_\infty$, where $x_a, x_{wm}$ $\in$ $[-1,1]^{128\times128}$
    \item SSIM score between attacked image and victim watermarked image, $\text{SSIM}(x_a,x_{wm})$
\end{enumerate}

\subsection{64 bit CNN Architecture \texttt{HiDDeN} Hyper-parameter Tuning}
\label{64bits_cnn_tuning}

As mentioned in \S~\ref{ss:models}, we perform grid search on \texttt{encoder\_loss} weight in $\{0.7, 0.35, 0.175, 0.0875\}$ and \texttt{decoder\_loss} weight in $\{1, 2, 4, 8\}$ when training target \texttt{HiDDeN} with CNN architecture and $\ell=64$ on DiffusionDB~\citep{wang2022diffusiondb} dataset. We choose the pair that gives the highest \texttt{BA} while having a mean square error less than $0.02$, both metrics measured on the validation set. The final chosen hyperparameters are \texttt{encoder\_loss} weight $0.7$ and decoder\_loss weight $4$. For the MidJourney~\citep{midjourney2022} dataset, we keep the same \texttt{encoder\_loss} and \texttt{decoder\_loss} weights.

\subsection{\text{MidJourney} Results for \S~\ref{sec:exp2}}
\label{appendix:mjresults}

Putting DiffusionDB~\citep{wang2022diffusiondb} and MidJourney~\citep{midjourney2022} results together, we can see OFT ($k=1$) is better than or comparable with~\citet{hu2024transfer} (the strongest one, $k=50$) in 11 out of 12 configurations except $\ell=20$ ResNet architecture on DiffusionDB~\citep{wang2022diffusiondb}. 

\begin{figure}[h!]
\begin{center}
\includegraphics[width=0.75\linewidth]{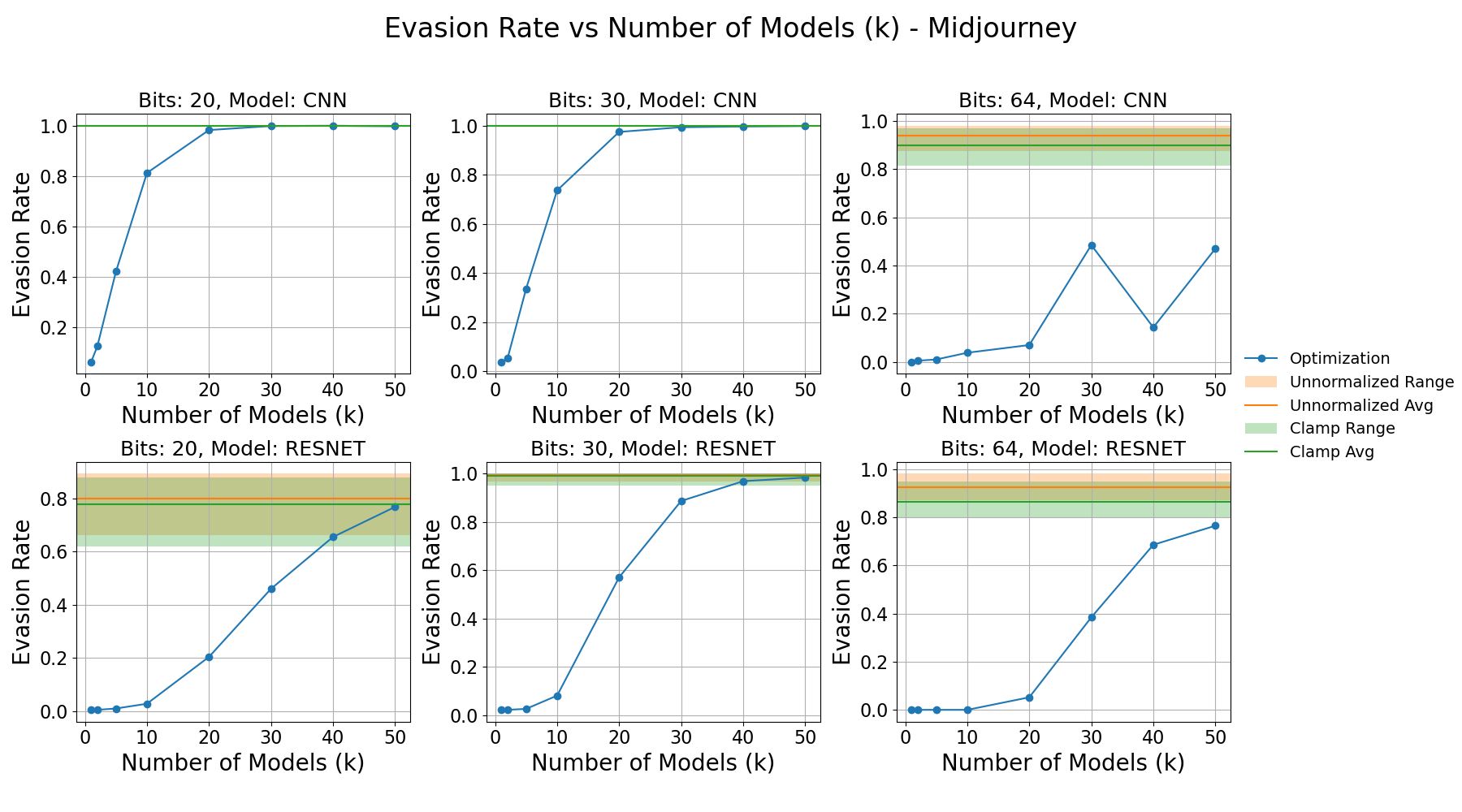} 
\end{center}
\caption{{\bf Evasion rate comparing~\citet{hu2024transfer} and OFT ($k=1$) on MidJourney:} Observe that OFT is superior most often, consistent with DiffusionDB results.}
\label{evasion_mj}
\end{figure}

\vspace{1mm} \noindent $\circ$ Figure~\ref{evasion_mj}: The Evasion Rate results on MidJourney are consistent with that for DiffusionDB. In all 6 configurations OFT ($k=1$) is comparable or better than \citet{hu2024transfer} ($k=50$). Another difference is that \citet{hu2024transfer} $k=40$ on $\ell=64$ CNN architecture has significantly lower evasion rate than $k=30$, showing potential instability of \citet{hu2024transfer}.

\begin{figure}[h!]
\begin{center}
\includegraphics[width=0.75\linewidth]{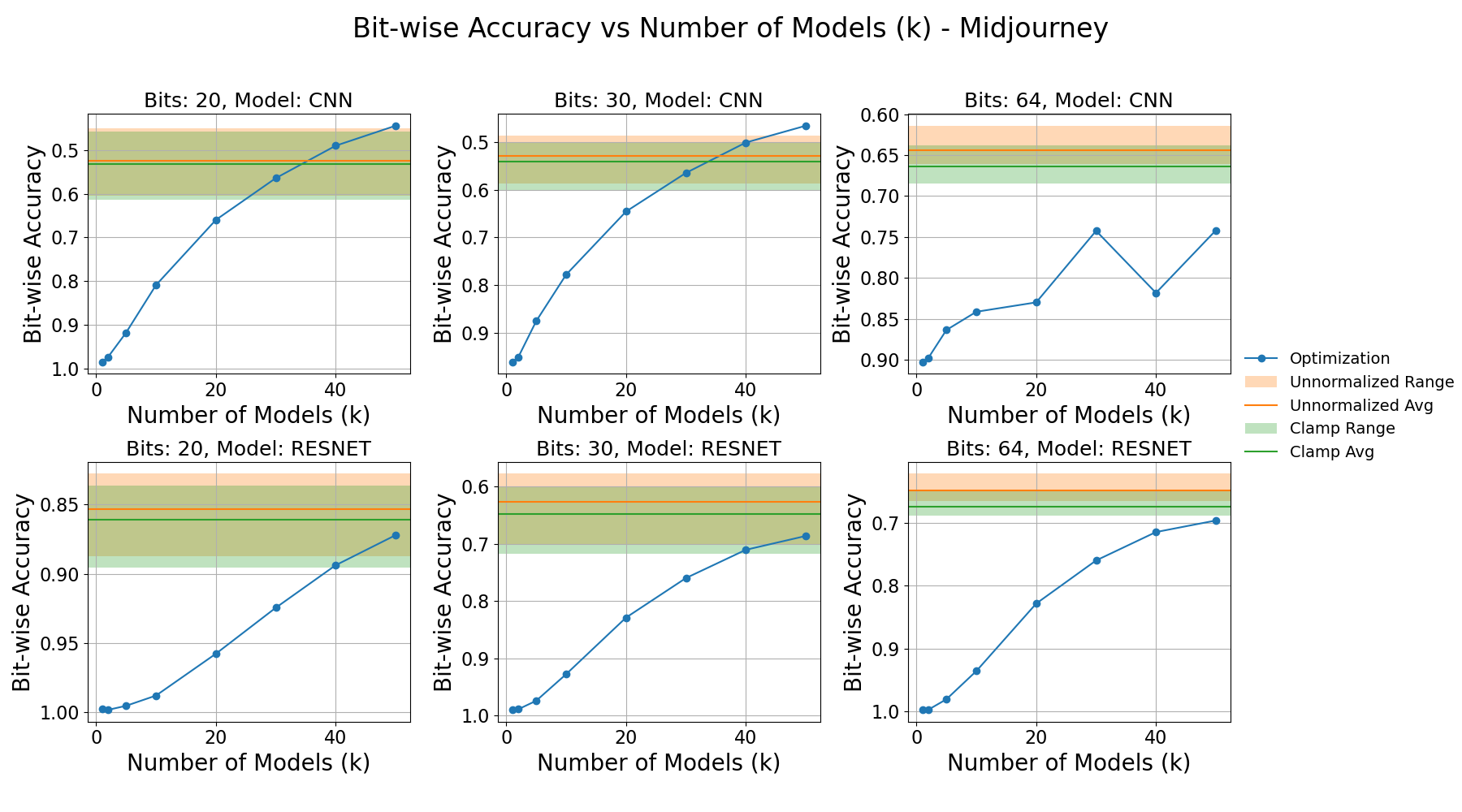} 
\end{center}
\caption{{\bf \texttt{BA} comparing~\citet{hu2024transfer} and OFT ($k=1$) on MidJourney:} Observe that OFT is superior most often, consistent with DiffusionDB results.}
\label{ba_mj}
\end{figure}

\vspace{1mm} \noindent $\circ$ Figure~\ref{ba_mj}: The \texttt{BA} results on MidJourney are consistent with that for DiffusionDB. Note that while for the CNN architecture and $\ell=20$, $\ell=30$,~\citet{hu2024transfer} has lower \texttt{BA} on $k=40$ and $k=50$, it is more than necessary: both OFT ($k=1$) and \citet{hu2024transfer} ($k=30, 40$) have saturated Evasion Rate at or very close to 1. Also, if \texttt{BA} is far below 0.5, then there are risks of being detected by the other end of two-tailed detection.

\begin{figure}[h!]
\begin{center}
\includegraphics[width=0.75\linewidth]{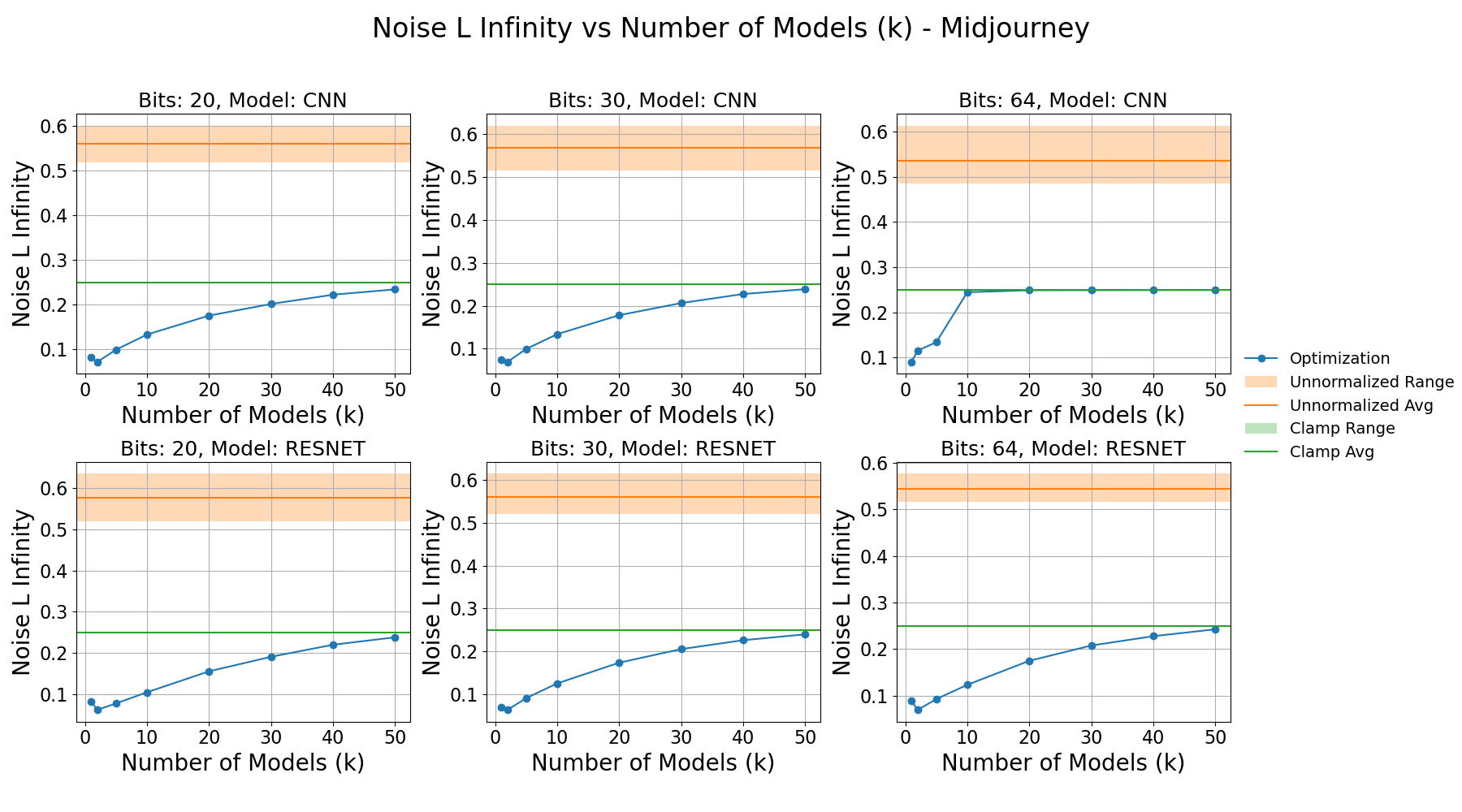} 
\end{center}
\caption{{\bf $\ell_\infty$ comparing~\citet{hu2024transfer} and OFT ($k=1$) on MidJourney:} Normalization (Clamp) ensures the noise is always within budget and significantly reduces $\ell_\infty$.}
\label{linf_mj}
\end{figure}

\vspace{1mm} \noindent $\circ$ Figure~\ref{linf_mj}: $\ell_\infty$ results on MidJourney are shown in Figure~\ref{linf_mj}. Normalization (Clamp) ensures the noise is always within budget and significantly reduces $\ell_\infty$. \citet{hu2024transfer} introduces perturbations with larger $\ell_\infty$ as $k$ gets larger in general.

\begin{figure}[h!]
\begin{center}
\includegraphics[width=0.75\linewidth]{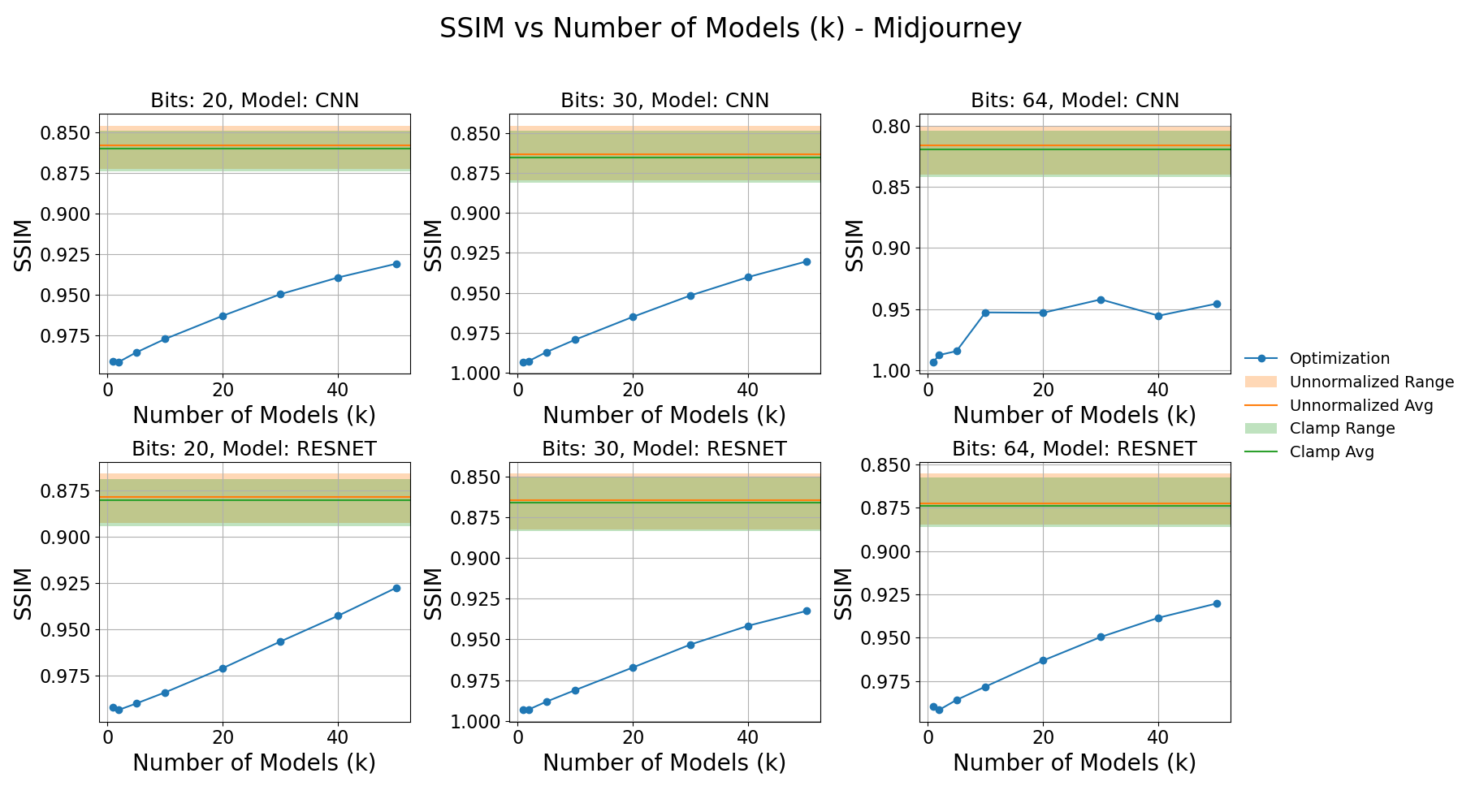} 
\end{center}
\caption{{\bf SSIM comparing~\citet{hu2024transfer} and OFT ($k=1$) on MidJourney:}~\citet{hu2024transfer} has better SSIM but larger $k$ harms SSIM in general; Normalization (Clamp) hardly helps improve SSIM.}
\label{ssim_mj}
\end{figure}

\vspace{1mm} \noindent $\circ$ Figure~\ref{ssim_mj}: SSIM results on MidJourney are shown in Figure~\ref{ssim_mj}. ~\citet{hu2024transfer} has better SSIM but larger $k$ harms SSIM in general; Normalization (Clamp) hardly helps improve SSIM.

\begin{figure}[h!]
\begin{center}
\includegraphics[width=0.75\linewidth]{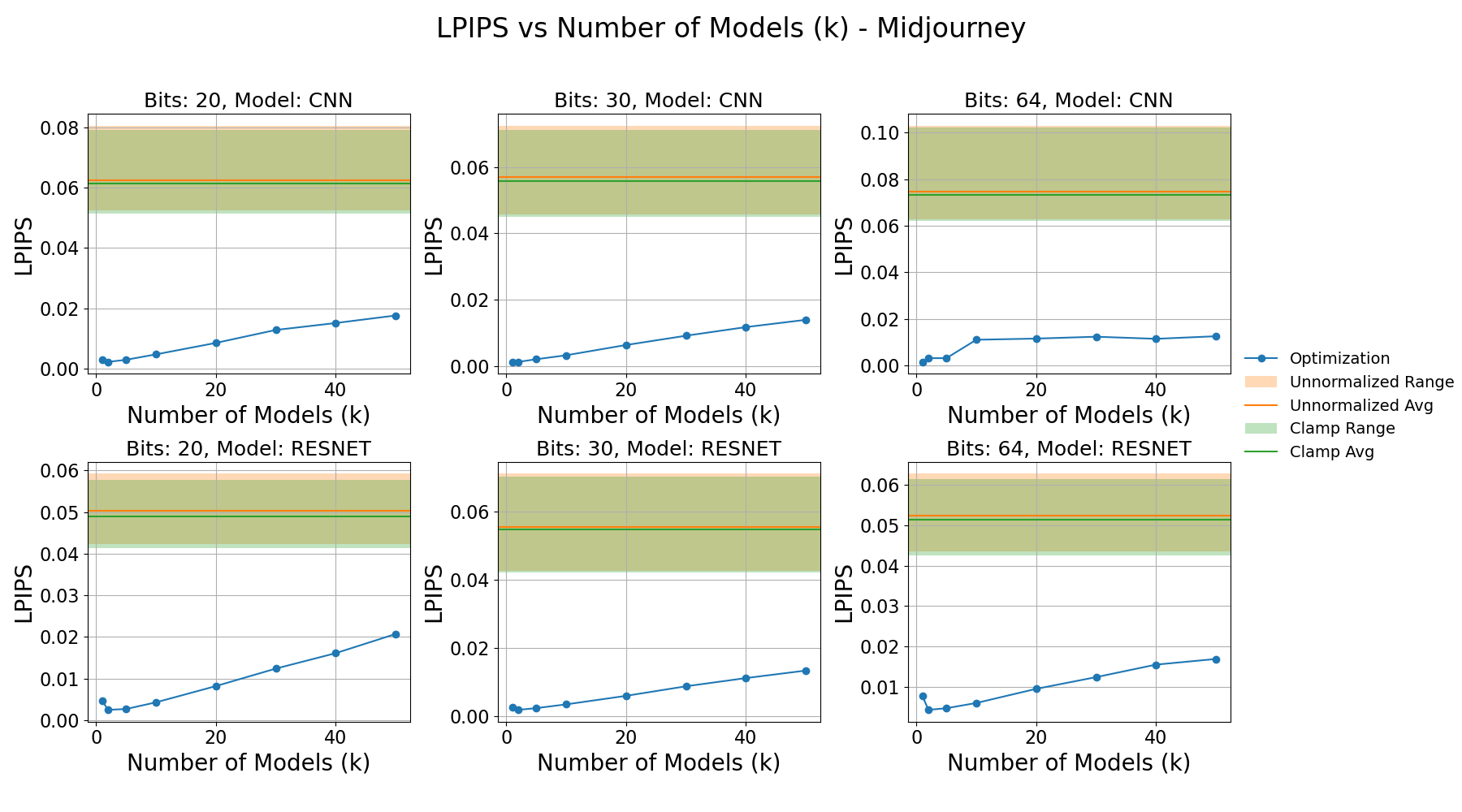} 
\end{center}
\caption{{\bf LPIPS comparing~\citet{hu2024transfer} and OFT ($k=1$) on MidJourney:}~\citet{hu2024transfer} has better LPIPS but larger $k$ harms LPIPS in general; Normalization (Clamp) hardly helps improve LPIPS.}
\label{lpips_mj}
\end{figure}

\vspace{1mm} \noindent $\circ$ Figure~\ref{lpips_mj}: LPIPS results on MidJourney are shown in Figure~\ref{lpips_mj}, which are consistent with SSIM results.

\FloatBarrier

\subsection{Additional DiffusionDB Results for \S~\ref{sec:exp2}}

\begin{figure}[h]
\begin{center}
\includegraphics[width=0.75\linewidth]{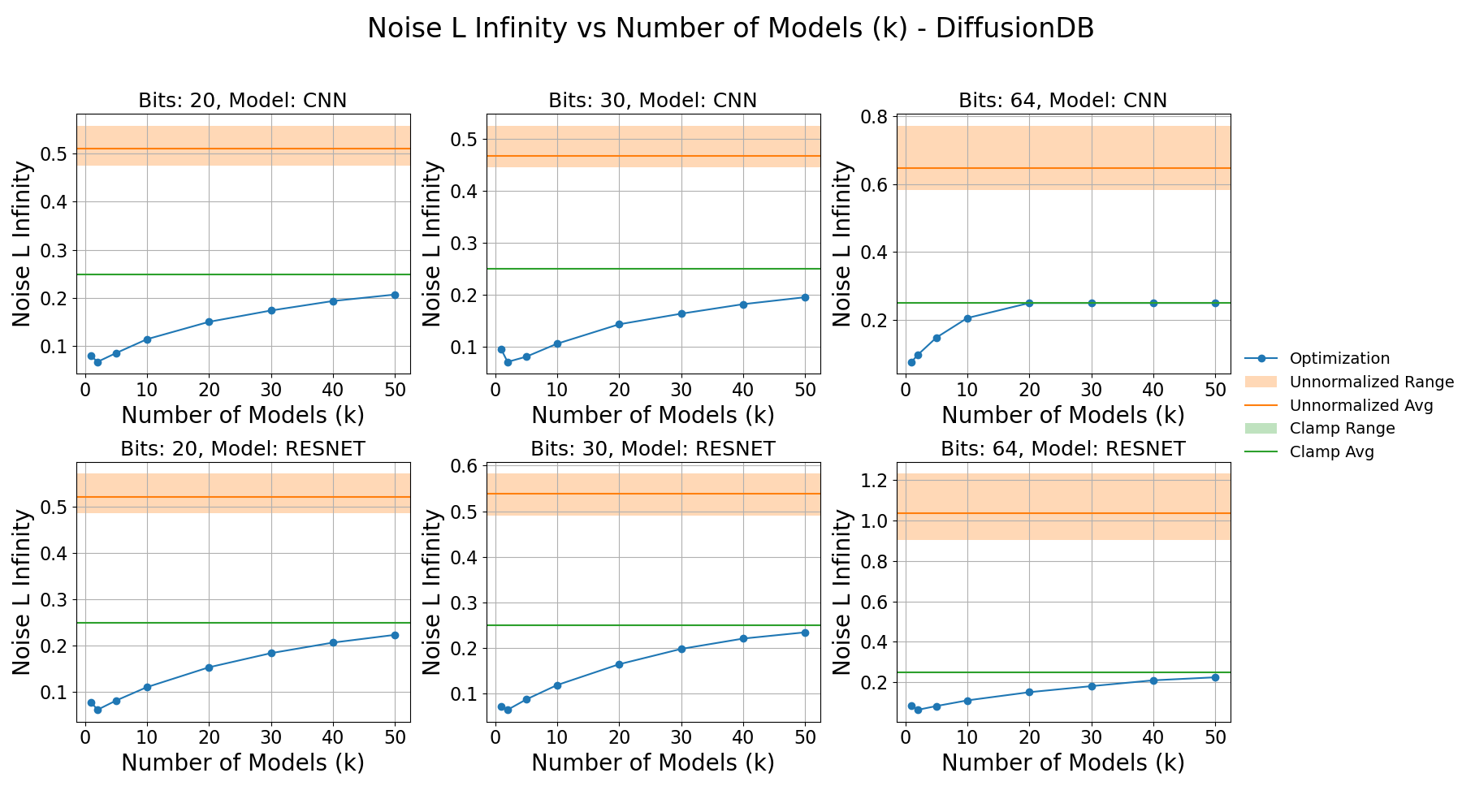} 
\end{center}
\caption{{\bf $\ell_\infty$ comparing~\citet{hu2024transfer} and OFT ($k=1$) on DiffusionDB:} Normalization (Clamp) ensures the noise is always within budget and significantly reduces $\ell_\infty$, consistent with MidJourney results.}
\label{linf_db}
\end{figure}

$\ell_\infty$ results on DiffusionDB are shown in Figure~\ref{linf_db}, which are consistent with MidJourney results.

\begin{figure}[h!]
\begin{center}
\includegraphics[width=0.75\linewidth]{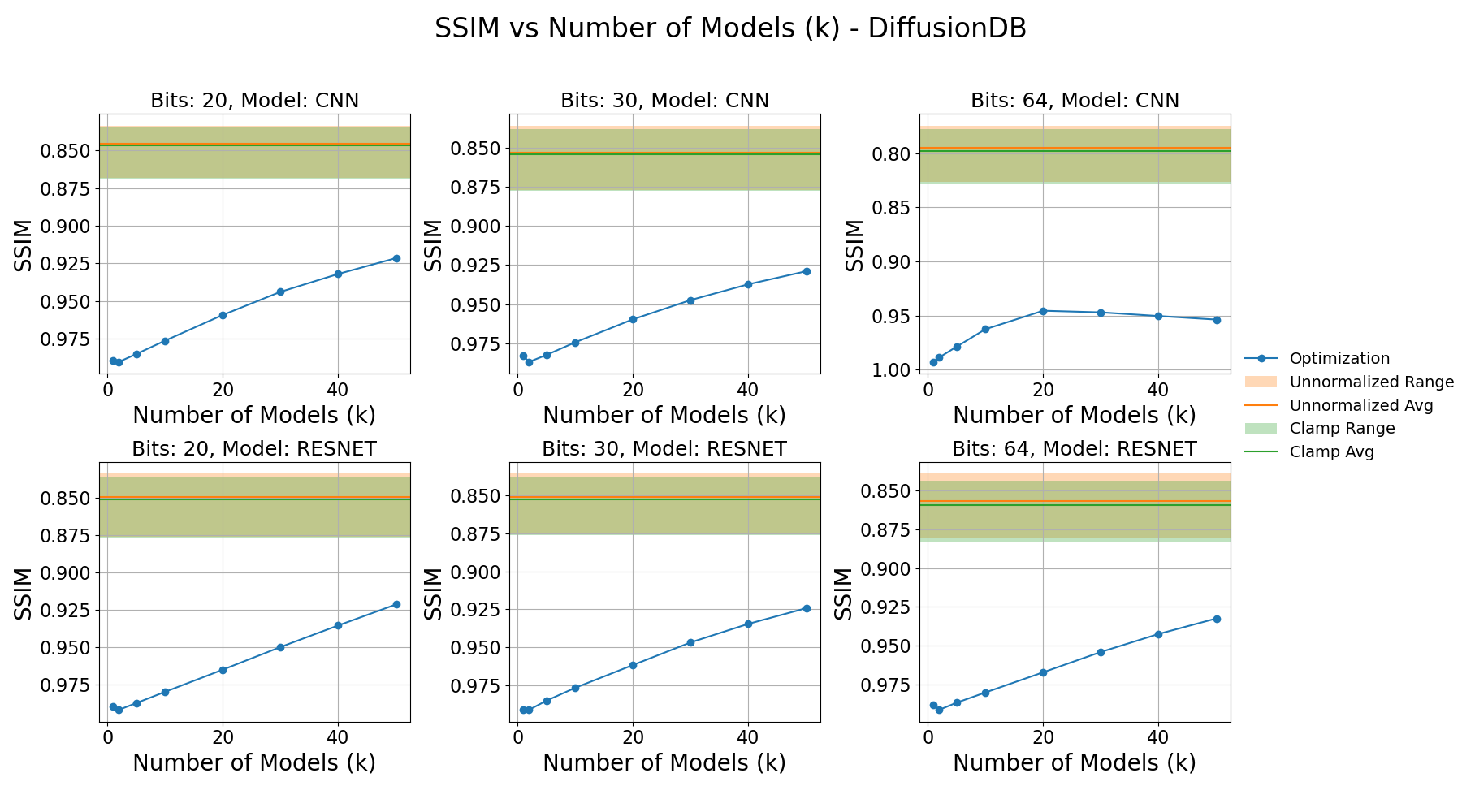} 
\end{center}
\caption{{\bf SSIM comparing~\citet{hu2024transfer} and OFT ($k=1$) on DiffusionDB:}~\citet{hu2024transfer} has better SSIM but larger $k$ harms SSIM in general; Normalization (Clamp) hardly helps improving SSIM. Results are consistent with MidJourney.}
\label{ssim_db}
\end{figure}

SSIM results on DiffusionDB are shown in Figure~\ref{ssim_db}, which are consistent with MidJourney results.

\begin{figure}[h!]
\begin{center}
\includegraphics[width=0.75\linewidth]{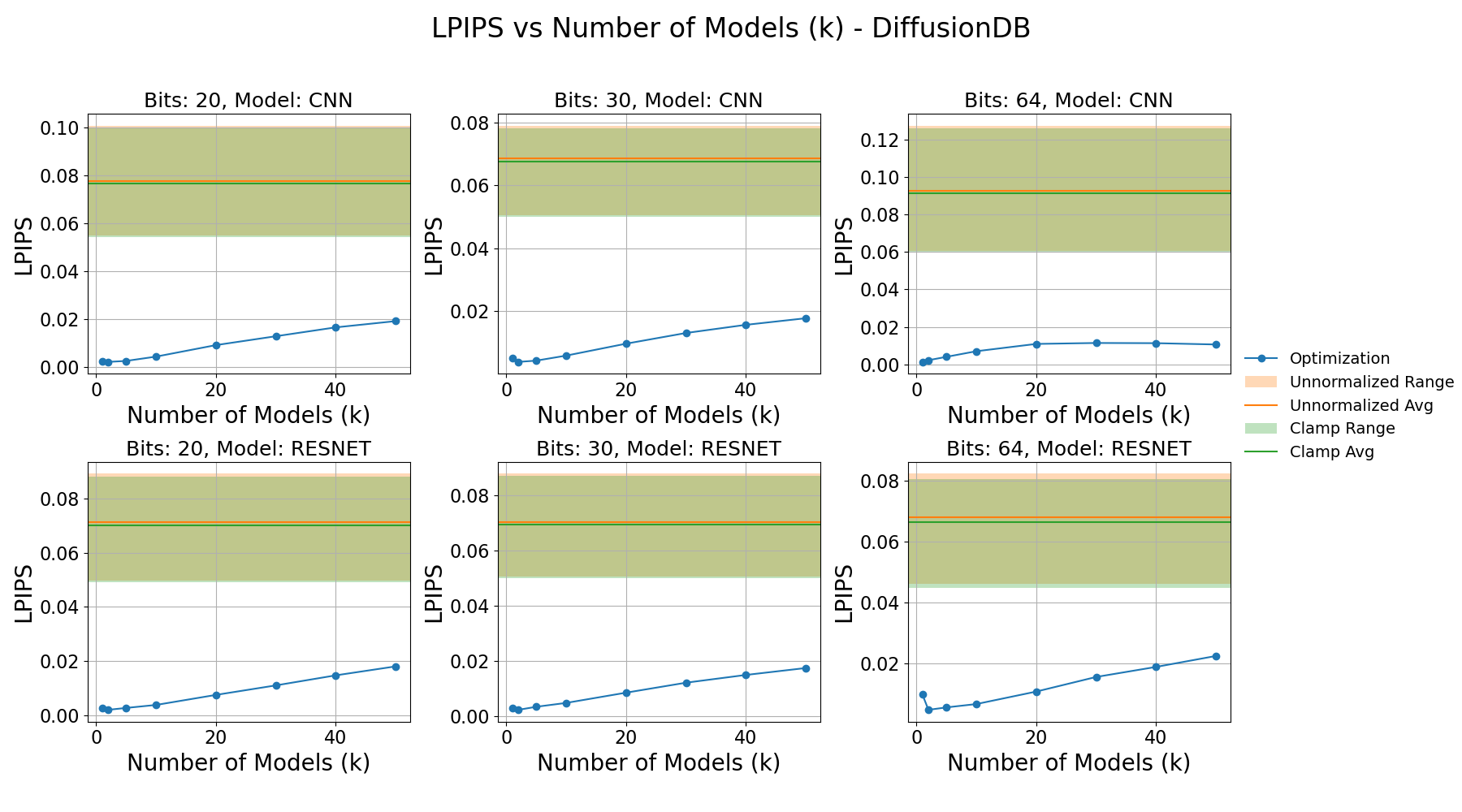} 
\end{center}
\caption{{\bf LPIPS comparing~\citet{hu2024transfer} and OFT ($k=1$) on DiffusionDB:}~\citet{hu2024transfer} has better LPIPS but larger $k$ harms LPIPS in general; Normalization (Clamp) hardly helps improving LPIPS. Results are consistent with MidJourney.}
\label{lpips_db}
\end{figure}

LPIPS results on DiffusionDB are shown in Figure~\ref{lpips_db}, which are consistent with MidJourney results.

\FloatBarrier

\subsection{Additional Results for \S~\ref{sec:exp1}}

Perceptibility results, including $\ell_\infty$, SSIM, and LPIPS, skipped in \S~\ref{sec:exp1} are not listed for the following reason: the only successful attack is when the target model's method matches the surrogate models' method (\texttt{HiDDeN}), so perceptibility results are not meaningful for failed attacks.

\begin{figure}[H]
\begin{subfigure}{0.48\columnwidth}
    \includegraphics[width=\linewidth]{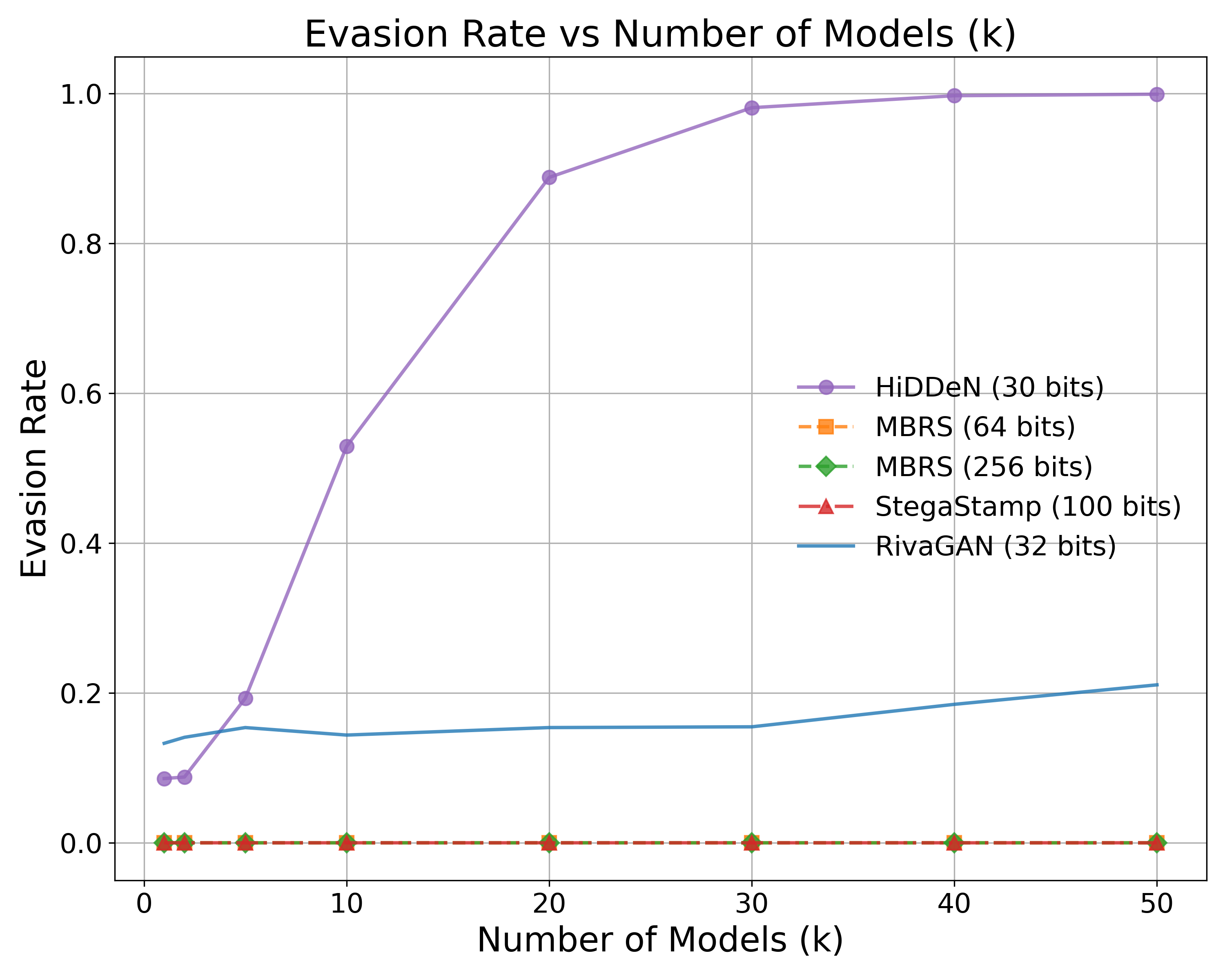}
    \caption{{\bf Evasion rate}}
\end{subfigure}
\hfill
\begin{subfigure}{0.48\columnwidth}
    \includegraphics[width=\linewidth]{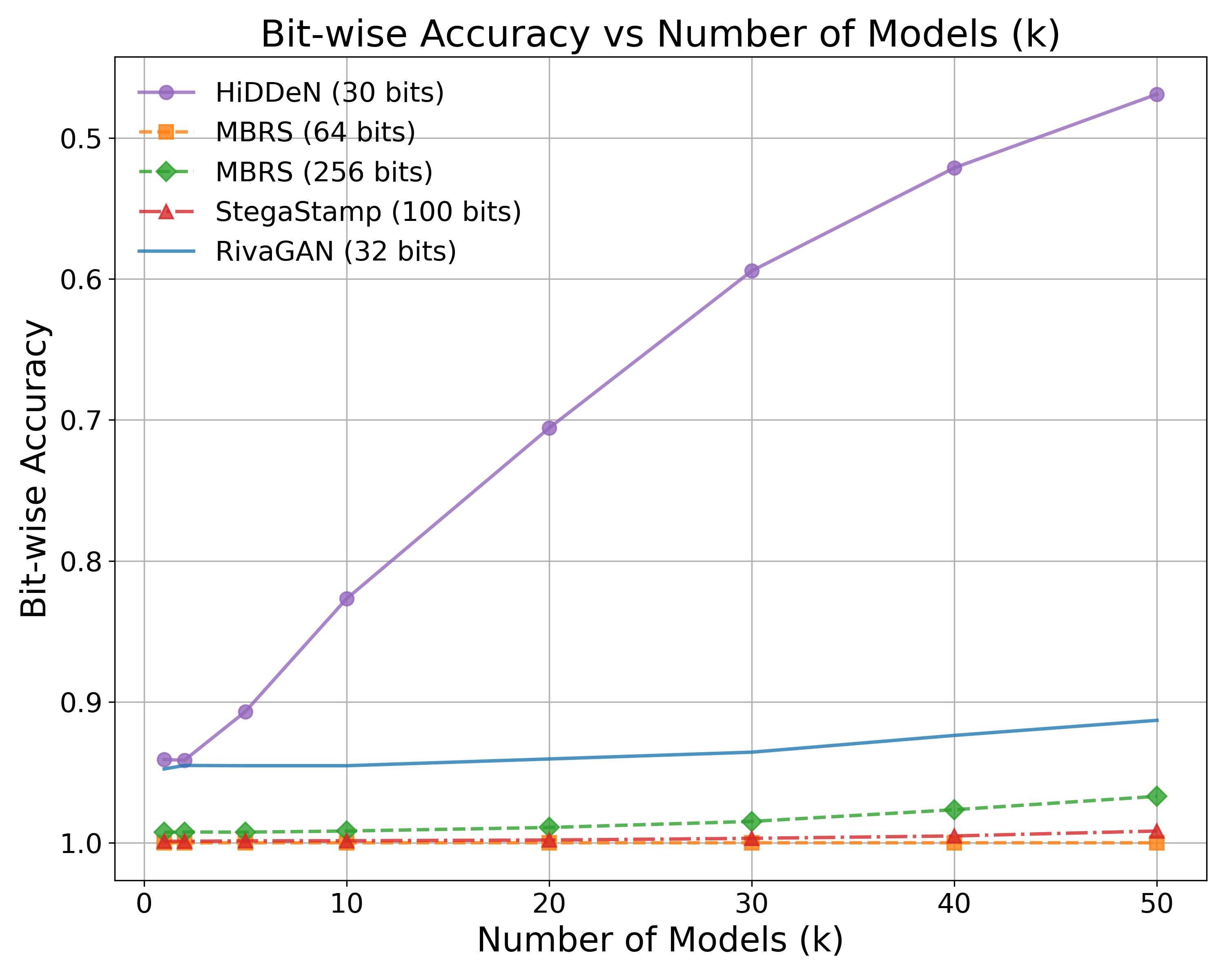}
    \caption{{\bf \texttt{BA}}}
\end{subfigure}
\caption{{\bf Evaluation of~\citet{hu2024transfer}'s attack to different watermarking methods, with the official unaligned \texttt{MBRS} (256 bits) added.} Still, the only successful attack is when the target model's method matches the surrogate models' method (\texttt{HiDDeN}).}
\label{exp_1_additional}
\end{figure}

We add the unaligned official \texttt{MBRS} ($\ell=256$) for the experiment in \S~\ref{sec:exp1} to be consistent with unaligned \texttt{StegaStamp} and \texttt{RivaGAN}. The Evasion Rate and \texttt{BA} result with the unaligned official \texttt{MBRS}, known as \texttt{MBRS} (256 bits), is shown in Figure~\ref{exp_1_additional}. The aligned \texttt{MBRS} is shown as \texttt{MBRS} (64 bits), whose configuration is detailed in Appendix~\ref{ss:mbrs} and the alignment is discussed in Appendix~\ref{sec:app_align}.
The unaligned \texttt{MBRS} results are consistent with that of the partially aligned version, i.e., no successful transfer, showing that the required aligning efforts for successful transfer are too high to be practical. 

\section{Naive Transfer}
\label{app:naive}

\begin{figure}[H]
\begin{subfigure}{0.48\columnwidth}
    \includegraphics[width=\linewidth]{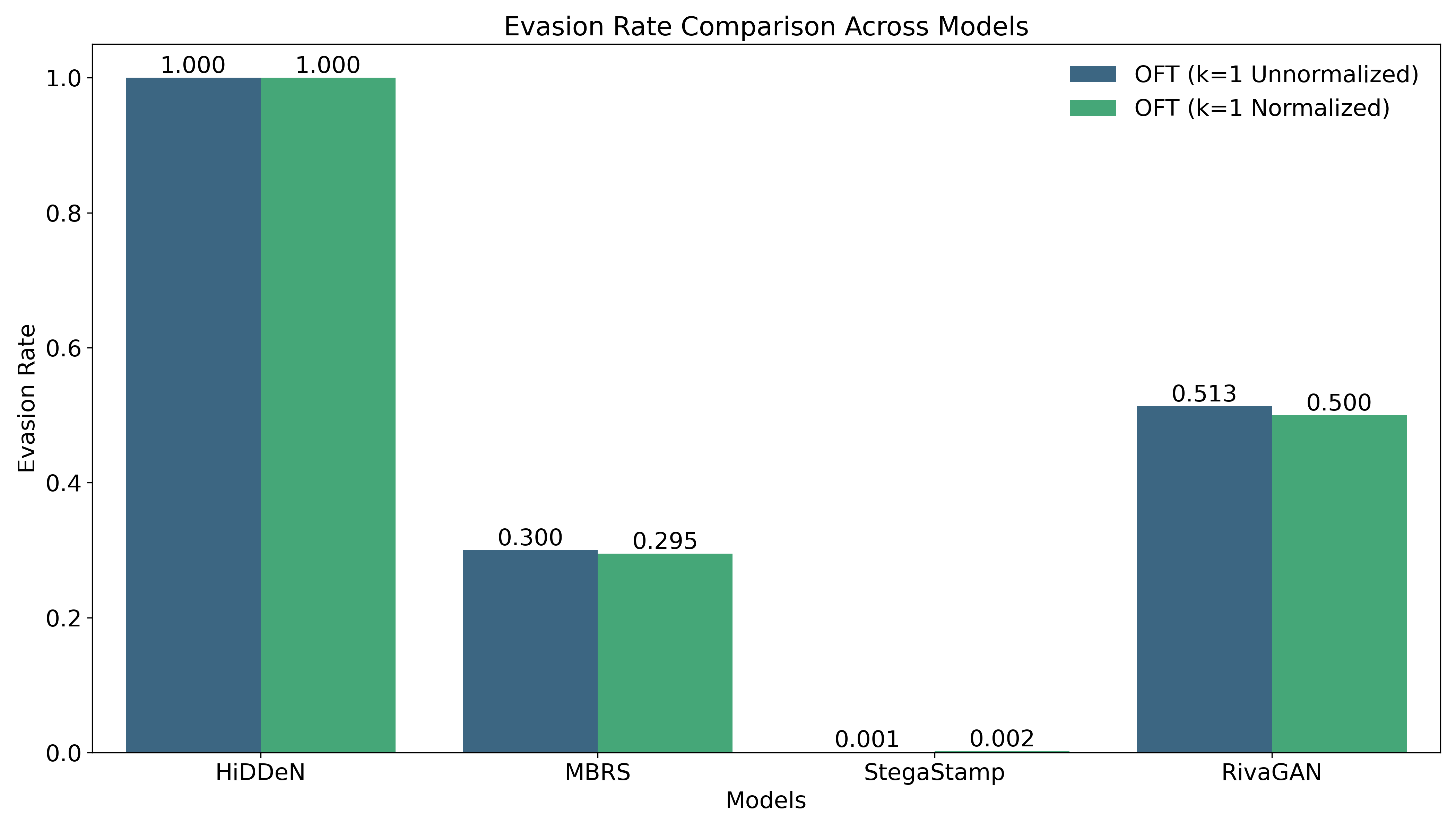}
    \caption{{\bf Evasion rate}}
\end{subfigure}
\hfill
\begin{subfigure}{0.48\columnwidth}
    \includegraphics[width=\linewidth]{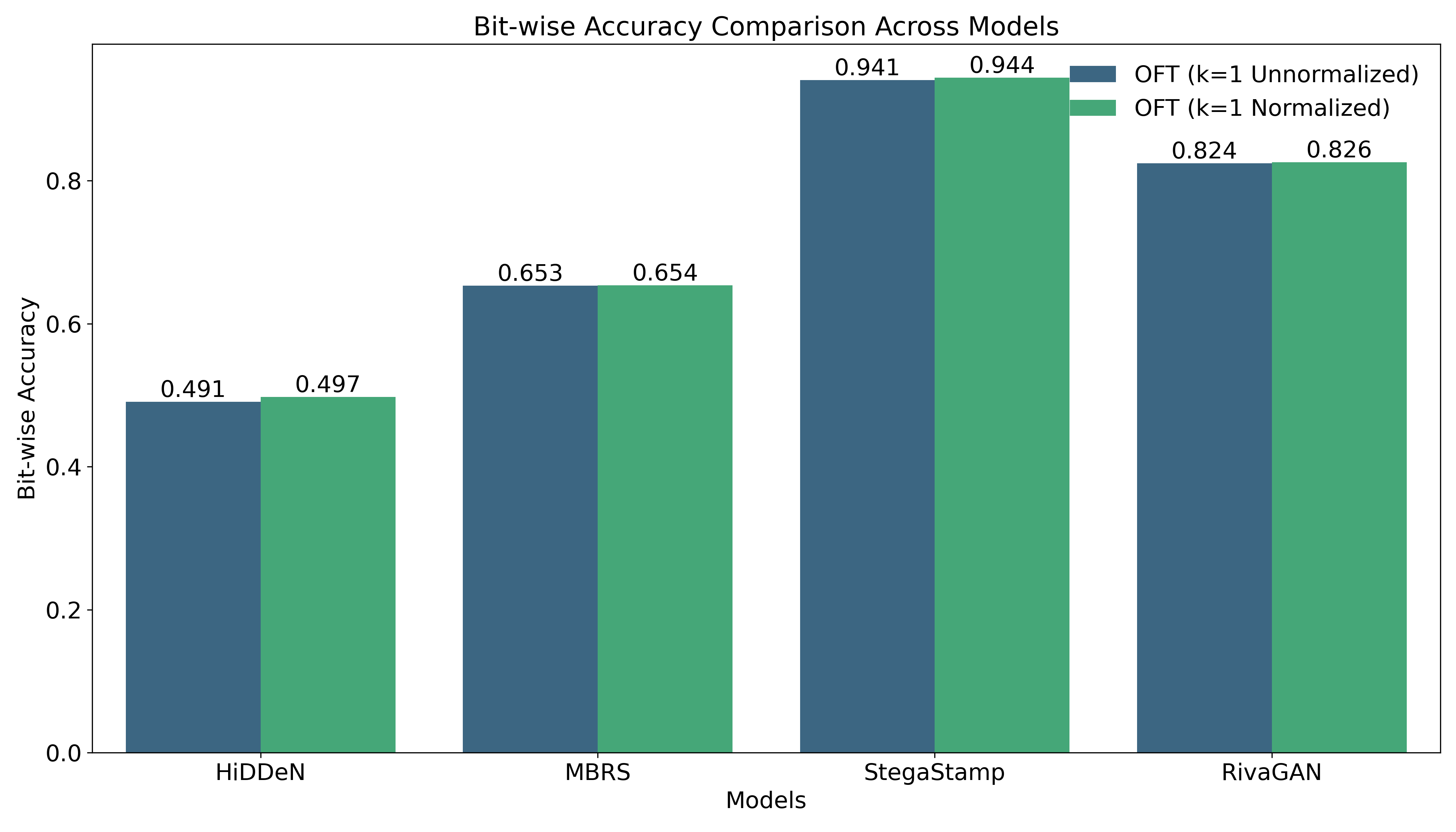}
    \caption{{\bf \texttt{BA}}}
\end{subfigure}
\caption{{\bf Evaluation of OFT ($k=1$) attack to different watermarking methods.} The only successful attack is when the target model's method matches the surrogate models' method (\texttt{HiDDeN}). While consistent with \citet{hu2024transfer} in \S\ref{sec:exp1} in general, OFT is relatively more effective.}
\label{naive}
\end{figure}

We perform a naive transfer attack using OFT on different target models with $k=1$. The target \texttt{HiDDeN} is trained on DiffusionDB, with default $\ell=30$ and CNN architecture. This reflects the case where configurations are not aligned and the attacker has limited computation capability. Results are shown in Figure~\ref{naive}. Only the control group, i.e., when the target model is the same as the surrogate model leads to a successful attack. This motivates us to design experiments relaxing one single assumption each time.

Note that while Table~\ref{comparison} shows \texttt{HiDDeN} and \texttt{StegaStamp} have similar encoder blocks, their encoders' architectures are actually very different. Firstly, as listed in Appendix~\ref{ss:hidden_arch} and \ref{ss:ss_arch}, most layers for the encoder \texttt{Enc} of \texttt{HiDDeN} are in $e$ before merging, while most layers for the encoder \texttt{Enc} of \texttt{StegaStamp} are in $h$ after merging. Also, U-Net style architecture includes many shortcuts, which are not in \texttt{HiDDeN}. Such architecture differences explain why \texttt{HiDDeN} transfers poorly to \texttt{StegaStamp} shown in Figure~\ref{naive}.

\section{Aligning MBRS}
\label{sec:app_align}
For the dataset, we use the same training set DALLE-2~\citep{dalle_gallery_2023} as surrogate models to train the target MBRS model. We replace the JPEG-related noise with the same noise as surrogate models during training. We also align the input image dimension from $256\times256$ to $128\times128$. We also try to change $\ell$ from $256$ to $30$ to align with the surrogate models, but find that it is impossible and end up with a compromising $\ell=64$. The secret message is reshaped as a two-dimensional square array and it goes through $4$ \texttt{ExpandNet} layers where each layer expands both dimensions by 2. Consequently, even if we change the number of \texttt{ExpandNet} layers, $\ell$ can only be within $\{4^0,4^1,4^2,4^3,4^4,...,4^7\}$. $\ell=4^2=16$ or fewer bits is too short for practice. Consequently, $\ell=4^3=64$ is chosen as it is the closest to $30$ in the remaining possible values and does not change the number of layers in \texttt{ExpandNet}.



\end{document}